\documentstyle[12pt,epsf]{article} \hoffset -0.5in \textwidth 6.00in
\textheight
8.5in  \parskip 7pt \openup1.5\jot \parindent=0.5in
\newskip\humongous \humongous=0pt plus 1000pt minus 1000pt
\def\caja{\mathsurround=0pt}
\def\eqalign#1{\,\vcenter{\openup1\jot \caja
        \ialign{\strut \hfil$\displaystyle{##}$&$
        \displaystyle{{}##}$\hfil\crcr#1\crcr}}\,}
\newif\ifdtup

\def\eqright #1\cr{\noalign{\hfill$\displaystyle{{}#1}$}}
\def\eqleft #1\cr{\noalign{\noindent$\displaystyle{{}#1}$\hfill}}

\def\oldreffmt#1{\rlap{[#1]} \hbox to 2\parindent{}}

\def\figfmt#1{\rlap{Figure {#1}} \hbox to 1in{}}

\def\VEV#1{\left\langle #1\right\rangle}

\def\sectioneq{\def\theequation{\thesection.\arabic{equation}}{\let
\holdsection=\section\def\section{\setcounter{equation}{0}\holdsection}}}%

\def\auto{\eqno(\refstepcounter{equation}\theequation)}
\def\begineq #1\endeq{$$ \refstepcounter{equation}\eqalign{#1}\eqno
	(\theequation) $$}
\def\contlimit{\,{\hbox{$\longrightarrow$}\kern-1.8em\lower1ex
\hbox{${\scriptstyle (a\rightarrow0)}$}}\,}
\def\centeron#1#2{{\setbox0=\hbox{#1}\setbox1=\hbox{#2}\ifdim
\wd1>\wd0\kern.5\wd1\kern-.5\wd0\fi
\copy0\kern-.5\wd0\kern-.5\wd1\copy1\ifdim\wd0>\wd1
\kern.5\wd0\kern-.5\wd1\fi}}
\def\centerover#1#2{\centeron{#1}{\setbox0=\hbox{#1}\setbox
1=\hbox{#2}\raise\ht0\hbox{\raise\dp1\hbox{\copy1}}}}
\def\centerunder#1#2{\centeron{#1}{\setbox0=\hbox{#1}\setbox
1=\hbox{#2}\lower\dp0\hbox{\lower\ht1\hbox{\copy1}}}}
\def\lsim{\;\centeron{\raise.35ex\hbox{$<$}}{\lower.65ex\hbox
{$\sim$}}\;}
\def\gsim{\;\centeron{\raise.35ex\hbox{$>$}}{\lower.65ex\hbox
{$\sim$}}\;}
\def\st#1{\centeron{$#1$}{$/$}}

\def\super#1{\ifmmode \hbox{\textsuper{#1}}\else\textsuper{#1}\fi}
\def\textsuper#1{\newcount\holdspacefactor\holdspacefactor=\spacefactor
$^{#1}$\spacefactor=\holdspacefactor}

\def\getcite#1,{\advance\citenumber by1
\def\getcitearg{#1}\def\lastarg{@}
\ifnum\citenumber=1
\ref{#1}\let\next=\getcite\else\ifx\getcitearg\lastarg\let\next=\relax
\else ,\ref{#1}\let\next=\getcite\fi\fi\next}

\def\pom{{\rm P\kern -0.53em\llap I\,}}
\def\spom{{\rm P\kern -0.36em\llap \small I\,}}
\def\sspom{{\rm P\kern -0.33em\llap \footnotesize I\,}}

\relax

\newskip\humongous \humongous=0pt plus 1000pt minus 1000pt
\def\caja{\mathsurround=0pt}
\def\eqalign#1{\,\vcenter{\openup1\jot \caja
        \ialign{\strut \hfil$\displaystyle{##}$&$
        \displaystyle{{}##}$\hfil\crcr#1\crcr}}\,}
\newif\ifdtup

\def\eqright #1\cr{\noalign{\hfill$\displaystyle{{}#1}$}}
\def\eqleft #1\cr{\noalign{\noindent$\displaystyle{{}#1}$\hfill}}

\def\oldreffmt#1{\rlap{[#1]} \hbox to 2\parindent{}}

\def\figfmt#1{\rlap{Figure {#1}} \hbox to 1in{}}

\def\VEV#1{\left\langle #1\right\rangle}

\def\auto{\eqno(\refstepcounter{equation}\theequation)}
\def\begineq #1\endeq{$$ \refstepcounter{equation}\eqalign{#1}\eqno
	(\theequation) $$}
\def\contlimit{\,{\hbox{$\longrightarrow$}\kern-1.8em\lower1ex
\hbox{${\scriptstyle (a\rightarrow0)}$}}\,}
\def\centeron#1#2{{\setbox0=\hbox{#1}\setbox1=\hbox{#2}\ifdim
\wd1>\wd0\kern.5\wd1\kern-.5\wd0\fi
\copy0\kern-.5\wd0\kern-.5\wd1\copy1\ifdim\wd0>\wd1
\kern.5\wd0\kern-.5\wd1\fi}}
\def\centerover#1#2{\centeron{#1}{\setbox0=\hbox{#1}\setbox
1=\hbox{#2}\raise\ht0\hbox{\raise\dp1\hbox{\copy1}}}}
\def\centerunder#1#2{\centeron{#1}{\setbox0=\hbox{#1}\setbox
1=\hbox{#2}\lower\dp0\hbox{\lower\ht1\hbox{\copy1}}}}
\def\lsim{\;\centeron{\raise.35ex\hbox{$<$}}{\lower.65ex\hbox
{$\sim$}}\;}
\def\gsim{\;\centeron{\raise.35ex\hbox{$>$}}{\lower.65ex\hbox
{$\sim$}}\;}
\def\st#1{\centeron{$#1$}{$/$}}

\def\super#1{\ifmmode \hbox{\textsuper{#1}}\else\textsuper{#1}\fi}
\def\textsuper#1{\newcount\holdspacefactor\holdspacefactor=\spacefactor
$^{#1}$\spacefactor=\holdspacefactor}

\def\getcite#1,{\advance\citenumber by1
\ifnum\citenumber=1
\ref{#1}\let\next=\getcite\else\ifx#1@\let\next=\relax
\else ,\ref{#1}\let\next=\getcite\fi\fi\next}

\def\upon #1/#2 {{\textstyle{#1\over #2}}}
\relax

\def\mainhead#1{\setcounter{equation}{0}\addtocounter{section}{1}
  \vbox{\begin{center}\large\bf #1\end{center}}\nobreak\par}
\sectioneq
\def\subhead#1{\bigskip\vbox{\noindent\bf #1}\nobreak\par}

\def\til#1{\centeron{\hbox{$#1$}}{\lower 2ex\hbox{$\char'176$}}}
\def\tild#1{\centeron{\hbox{$\,#1$}}{\lower 2.5ex\hbox{$\char'176$}}}
\def\sumtil{\centeron{\hbox{$\displaystyle\sum$}}{\lower
-1.5ex\hbox{$\widetilde{\phantom{xx}}$}}}

\def\pom{{\rm P\kern -0.53em\llap I\,}}
\def\spom{{\rm P\kern -0.36em\llap \small I\,}}
\def\sspom{{\rm P\kern -0.33em\llap \footnotesize I\,}}

\newcommand{\bit}{\begin{itemize}}
\newcommand{\eit}{\end{itemize}}

\newcommand{\beq}{\begin{equation}}
\newcommand{\eeq}{\end{equation}}
\newcommand{\beqa}{\begin{eqnarray}}
\newcommand{\eeqa}{\end{eqnarray}}

\begin{document}

\begin{titlepage}

\rightline{\vbox{\halign{&#\hfil\cr
&ANL-HEP-PR-95-19\cr
&UF-IFT-HEP-95-21\cr}}}

\vspace{.4in}

\begin{center}

{\bf GAUGE THEORY HIGH-ENERGY BEHAVIOR FROM J-PLANE UNITARITY}
\footnote{Work supported by the U.S. Department of Energy, Division of High
Energy Physics, \newline Contracts W-31-109-ENG-38 and DEFG05-86-ER-40272}

\medskip

{Claudio Corian\`{o}$^{a,b}$\footnote{
coriano@phys.ufl.edu ~$^{\#}$arw@hep.anl.gov}
and \ Alan. R. White$^{a\#}$}

\vskip 0.6cm

\centerline{$^a$High Energy Physics Division}
\centerline{Argonne National Laboratory}
\centerline{9700 South Cass, Il 60439, USA.}
\vspace{0.5cm}

\centerline{$^b$Institute for Fundamental Theory}
\centerline{Department of Physics}
\centerline{ University of Florida at Gainesville, FL 32611, USA}
\vspace{0.5cm}

\end{center}

\begin{abstract}

In a non-abelian gauge theory the $t$-channel multiparticle unitarity
equations continued in the complex j-plane can be systematically expanded
around $j=1$. The combination of Ward identity constraints with unitarity is
sufficient to produce directly many results obtained by Regge limit
leading-log and next-to-leading log momentum space calculations. The $O(g^2)$
BFKL kernel is completely determined. $O(g^4)$ contributions to the kernel
are also determined, including the leading contribution of a new
partial-wave amplitude - previously identified as a separate forward
component with a holomorphically factorizable spectrum. For this amplitude
the only scale ambiguity is the overall normalization and it is anticipated
to be a new conformally invariant kernel. The results suggest that all
conformally invariant reggeon interactions are determined by $t$-channel
unitarity.

\end{abstract}

\end{titlepage}

\mainhead{1. INTRODUCTION}

The Regge limit of $QCD$ is a challenging theoretical problem which will
surely continue to be addressed for some time to come - particularly if the
solution is fundamental to obtaining a full solution of the
theory\cite{arw2,arw3}. The inter-relation of the Regge limit with the
small-x behavior of structure functions and other new ``hard diffractive''
experimental phenomena recently observed at HERA and the Tevatron Collider
has been the subject of much recent investigation. Indeed, the leading-log
BFKL equation\cite{bfkl},originally derived in the Regge limit, has now been
applied extensively in the study of parton distributions at small-x. A
further factor in stimulating recent theoretical activity has been the
development of new techniques which  offer the hope of greater insight into
the physics involved. In particular, a two-dimensional effective
theory\cite{kls,lip1} has been developed which reproduces the leading order
perturbative calculations\cite{bfkl,cl} and which it is hoped can be used as
input to some form of ``$s$-channel'' unitarisation scheme.

In this paper our emphasis will be on the development of new techniques for
the exploitation of ``$t$-channel'' unitarity, rather than $s$-channel
unitarity, to determine corrections to leading-order results. The kernel of the
BFKL equation can be viewed as a  2-2 reggeon interaction and in \cite{ker}
we have suggested that the non-leading, $O(g^4)$, ``scale-invariant'' part of
this kernel can be derived directly by a ``reggeon diagram technique''. This
technique assumes that reggeized gluon interactions can be constructed from
reggeon diagrams in which the gluon couples via a ``nonsense-zero'' triple
vertex. Effectively, all contributions of gluons to reggeon interactions are
assumed to be traceable back to $t$-channel reggeized gluon exchanges. The
justification for the use of reggeon diagrams in this manner is that
constraints of multiparticle $t$-channel unitarity are, at least partially,
satisfied. However, it is easy to criticize the assumptions made and our
aim in this paper is to give a more fundamental derivation of the results
directly from $t$-channel unitarity continued in the complex angular
momentum plane (the $j$-plane).

Our analysis is based entirely on combining unitarity and gauge invariance
with a weak-coupling expansion of the theory around $j=1$. We avoid momentum
space calculations altogether. It might be thought that some calculations are
needed to input the defining lagrangian. In fact, as we already pointed out in
\cite{ker}, gauge invariance can be input via the Ward identity requirement
that reggeon amplitudes vanish at zero tranverse momentum and the gauge
group can be inserted via the group structure of the lowest-order reggeon
interaction (i.e. the triple Regge vertex). We will obtain both the
reggeization and explicit higher-order results using only these ingredients
as defining elements of the theory. Consequently, we make at least partial
progress towards the direct perturbative construction of ``Yang-Mills
reggeon theories'' envisaged in \cite{ker}.

Perturbatively expanding the unitarity equations around $j=1$ should be
equivalent to expanding in powers of leading, next-to-leading etc., Regge
limit logarithms. This equivalence, at first sight, should enable us
to compare our $j$-plane analysis directly with momentum space calculations.
However, the unitarity analysis gives only the $t$-channel discontinuities
responsible for the leading infra-red behavior of amplitudes. Therefore
our formalism gives reliable results only at small transverse momentum.
Small has, of course, to be defined in terms of some scale which breaks
transverse momentum scale-invariance and we will not discuss
this. However, we expect that the infra-red analysis will be sufficient to
find all conformally invariant reggeon interactions and our results are
consistent with this expectation. (We should note that it is not yet clear
to what extent the momentum space evaluation\cite{fl} of non-leading log
amplitudes avoids ambiguities associated with the introduction of scales,
particularly that associated with renormalization and the large momentum
evolution of the coupling constant.)

A major issue is that the reggeon diagram method used in \cite{ker} implicitly
assumes a reduction to transverse momentum integrals which a-priori is not
justified and which next-to-leading order $s$-channel calculations
apparently do not give\cite{lip1}. We will show that this reduction is
always justified when only ``nonsense'' $t$-channel gluon states are
involved in producing a reggeon interaction. (In general, ``nonsense
states'' have less angular momentum than helicity - as a result of analytic
continuation in $j$. Note also that in our analysis the two-dimensional
``transverse momentum'' variables, that we refer to throughout, are the
$t$-channel timelike counterpart of $s$-channel transverse momenta.) We will
show that the BFKL kernel actually arises entirely from nonsense states, as
does that part of the $O(g^4)$ kernel that we have separated in \cite{cw} as
having distinct infra-red finiteness and holomorphic factorization
properties. Indeed we will show that this $O(g^4)$ contribution is actually
{\it a new partial-wave amplitude which appears for the first time at this
order.} Correspondingly the only scale ambiguity is the overall
normalization. In a companion paper\cite{cpw} we construct what we
conjecture to be the non-forward conformally invariant form of this
amplitude.

The complete $O(g^4)$ kernel that we gave in \cite{ker} can not be
unambiguously
derived from nonsense contributions. Even though we expect this kernel to be
an infra-red ``scale-invariant'' approximation to the complete kernel. We will
make very little reference to scale-dependent contributions. Note, however,
that in a recent paper Kisrchner\cite{kir} has discussed how the $O(g^4)$
kernel can arise as an approximation when the leading-log $s$-channel
multi-Regge effective lagrangian is used. A number of $s$-channel
contributions have to be combined to reproduce the simplicity of the
$t$-channel reggeon diagram results.

Even if the program\cite{fl} to calculate the next-to-leading order BFKL kernel
can be completed in momentum space, it seems inevitable that study of the
high-energy behavior of QCD will eventually move, in large part, to the
complex j-plane. The direct calculation of logarithms is extremely
complicated compared to the simplicity of the results when expressed in
j-plane language. (To realize the economy of language, one has only to
compare the simple Regge pole formula for electron exchange that is the
outcome, with the 240 pages of the Physical Review used\cite{mw} by McCoy
and Wu to calculate to twelfth order in QED.) All of $t$-channel unitarity,
particularly reggeon unitarity, can be simply expressed in the $j$-plane.
The structure of multiparticle partial-wave amplitudes is also only apparent
in this language. In particular the new $O(g^4)$ amplitude we discuss would
be very hard to isolate in $s$-channel calculations. In general it seems
likely that the complexity of an $s$-channel effective lagrangian
formalism\cite{lip1} will obscure many of the $t$-channel simplifications.
If the problem of higher-order corrections can be transferred to the
j-plane, as we will partially succeed in doing, then it is possible that
significantly more progress can be made. To put our efforts in context we
give a very brief historical review of the development of $t$-channel
unitarity as a tool to study both abstract Regge theory and the Regge
behavior of gauge theories in particular.

Reggeon diagrams, or Reggeon Field Theory (RFT) as the formalism came to be
called, originally arose from the ``hybrid Feynman diagram'' formalism of
Gribov\cite{gri}. Part of Gribov's motivation for developing the diagrammatic
formalism was to provide an interpretation of the abstract results of
Gribov, Pomeranchuk and Ter-Martirosyan (GPT) obtained from multiparticle
$t$-channel unitarity continued in the complex j-plane\cite{gpt}. The GPT work
was in turn a response to Mandelstam's work\cite{sm} showing that $t$-channel
unitarity could be reliably used to calculate the high-energy behaviour of
Feynman graphs giving Regge cut behaviour, whereas $s$-channel unitarity was
unreliable. (As remains true in current gauge theory calculations, there are
many cancellations amongst $s$-channel states, while $t$-channel states give
easily distinguished contributions.) The GPT results gave discontinuity
formulae for the angular momentum plane branch points (Regge cuts) due to
general multiple Regge pole exchange. The relationship of the reggeon
diagrams formulated by Gribov to the angular momentum plane unitarity
formulae was analagous to that of conventional Feynman diagrams to normal
momentum space unitarity. However, there were various analytic continuation
and summation ambiguities in the GPT formulae and, at the time, it seemed
that explicit diagrammatic calculations of the kind developed by Gribov
would be the only way to resolve such ambiguities. The direct diagrammatic
approach also seemed simpler than the complicated GPT formalism.

Subsequently major progress was made in understanding the analyticity
properties of multiparticle amplitudes\cite{arw1,harw,hs}. Asymptotic
dispersion relations were derived and provided the basis for a
comprehensive development of abstract multiparticle complex angular momentum
theory. All the ambiguities of the GPT work were resolved and the
resulting discontinuity formulae, or {\it reggeon unitarity equations}
were established. The most immediate application was to provide an
underlying framework for the study of Pomeron RFT. It was understood that,
given the lowest order reggeon interactions, the reggeon unitarity equations
determine the form of the general reggeon diagrams of a complete RFT. (The
generality of the circumstances under which the very attractive {\it
Critical Pomeron} solution\cite{cri} of RFT could appear as the true
asymptotic behavior of the strong interaction became particulary clear.)

The reggeization of the gluon, in a gauge theory, was actually first
obtained\cite{gs} by $j$-plane analysis of two-particle $t$-channel
unitarity. Only later was this confirmed in leading-log momentum-space
calculations. Once gluon reggeization is established, the reggeon unitarity
equations determine that the non-leading logs must reduce to an effective
two-dimensional theory which can be described in terms of reggeon
diagrams. Indeed, it has already been demonstrated\cite{reg} that in
Yang-Mills theories leading and next-to-leading logarithms are very
compactly described by reggeon diagrams - providing the most direct way to
derive the BFKL equation. We expect that a full set of such diagrams should
describe the leading power {\it plus all logarithms} obtained from
perturbation theory. In principle new Regge trajectories may emerge in
higher-orders, the symmetric octet trajectory found by Bartels and
W\"usthoff\cite{bw} being an example. However, although we will not elaborate
on our reasons, we do not expect this to be an
extended phenomenon.

At first sight only the general form of the phase-space in reggeon diagrams is
determined by reggeon unitarity and so it would be anticipated that
reggeon interactions should be
extracted from momentum space calculations as in \cite{reg}. A related point is
that, although reggeization was derived using two-particle unitarity\cite{gs},
it was necessary to first calculate the lowest-order nonsense amplitudes via
momentum space. Effectively, our purpose in this paper is to show that, when
gauge invariance and the gauge group are input as we have discussed,
reggeon unitarity can also be used to determine reggeon interactions in
gauge theories. It will be essential to expand the concept of reggeon
unitarity  to include the contribution of ``right-signature'' reggeon
nonsense states. The key to this will be, as we have already emphasized, the
weak-coupling expansion around $j=1$ - which is a ``nonsense point'' in a
reggeized vector theory. Since the reggeons of a gauge theory are
odd-signature reggeized gluons, the reggeon propagators and
interactions appearing in the diagrams contain particle poles in addition to
Regge poles. The particle poles give ``right-signature nonsense state
thresholds'' in $t$ (in contrast to the Regge cuts which can be described as
``wrong-signature nonsense state thresholds'' in $j$). It is by determining
the discontinuity across the $t$-thresholds directly from unitarity that we
actually determine the reggeon interactions.

We should note that the particle poles in reggeon diagrams also produce
infra-red divergences and as a consequence, the diagrams are
all infra-red divergent. We have discussed the dynamical implications of
these divergences, for the ``non-perturbative'' solution of the theory, at
length in \cite{arw2} and, of course, the divergences cancel in the physical
kernels appearing in the BFKL equation. In this paper we will regard the
divergences as a purely technical problem that we will ignore in deriving
explicit formulae. The massless theory is particularly simple and our
central purpose is to show how far we can go in this case.

Since the formalism we will use is very unfamiliar to most physicists we will
spend a considerable time just introducing concepts and language. We begin
in Section 2 by providing a very elementary outline of our analysis. In
Section 3 we cover various preliminary topics. We formulate the
leading-order BFKL equation in reggeon language, discuss the origin of
reggeon Ward identity constraints and introduce a diagrammatic notation for
color factors. Section 4 contains a review of nonsense states, transverse
momentum diagrams and Regge cut discontinuity formulae, as well as a
rederivation of the reggeization of the gluon in our formalism. In Section 5
we rederive the BFKL kernel before proceeding, in Section 6, to the
derivation of $O(g^4)$ reggeon interactions. The crucial result is the
demonstration that amongst the $O(g^4)$ interactions is a leading-order
contribution to a new partial-wave amplitude. Section 7 contains some
brief conclusions and comments.

\mainhead{2. OUTLINE OF THE ANALYSIS}

In this Section we outline the arguments developed at length in the
following Sections. We will avoid precise definitions in the interests of
presenting a simple overview.

We consider a reggeon theory containing a vector particle (which is, of
course, the gluon) lying on a Regge trajectory
$$
j~\equiv ~1~+~\omega~~=~\alpha(t)~= ~1~+~\Delta(q^2)~,
\auto\label{Del}
$$
which, if the particle is massless, satisfies
$$
\Delta(0)~=~\alpha(0)~- ~1~=~0
\auto\label{mless}
$$
In the $j$-plane there is a Regge pole - produced by a ``reggeon
propagator'' which we represent diagrammatically as in Fig.~2.1

\begin{center}
\leavevmode
\epsfxsize=2in
\epsffile{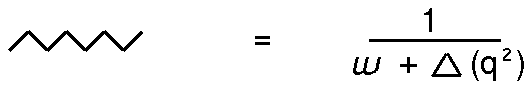}

Fig.~2.1 The Reggeon Propagator
\end{center}
As we review in Section 4, it follows from $t$-channel unitarity that there
will also be Regge cuts, or ``thresholds'', in the $j$-plane, arising from the
exchange of any number of reggeons. The $N$-reggeon cut arises from
phase-space integration of the $N$-reggeon propagator which, in the language
we develop, is simply a ``nonsense pole''. The phase-space is an integral
over the transverse momenta of the reggeons. (As we noted in the
Introduction, these are actually time-like two dimensional momenta in our
analysis). For example, the two-reggeon cut arises from
$$
\int {d^2k_1 \over k_1^2} {d^2k_2 \over k_2^2} \delta^2(q-k_1-k_2)~\Gamma_2
\auto\label{2rc}
$$
where $\Gamma_2$ is the two-reggeon propagator represented in Fig.~2.2.

\begin{center}
\leavevmode
\epsfxsize=4in
\epsffile{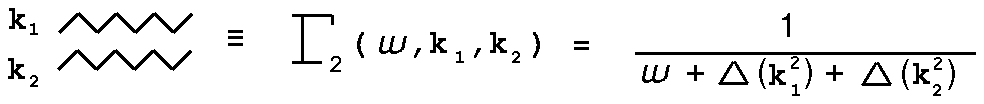}

Fig.~2.2 The two reggeon propagator
\end{center}
A complete set of reggeon diagrams satisfying reggeon unitarity is generated
by introducing a general set of reggeon interaction vertices coupling the
reggeon propagators. The lowest-order 2-2 reggeon vertex is the BFKL kernel.

We construct both the reggeon trajectory function and the reggeon
interaction vertices from the following ingredients.

\begin{description}

\item[{[2A]}] Gauge invariance is input via the Ward identity constraint that
all reggeon interaction vertices vanish when any reggeon transverse momentum
goes to zero.

\item[{[2B]}] The ``nonsense'' zero/pole structure required by general
analyticity properties is imposed, in addition to Ward Identity zeroes.

\item[{[2C]}] The group structure is input via the triple reggeon vertex,
which couples (a nonsense state of) two reggeons to a single reggeon carrying
transverse momentum $q$, i.e.

\begin{center}
\leavevmode
\epsfxsize=3in
\epsffile{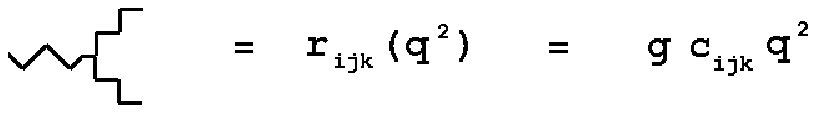}

\end{center}
$g$ is a dimensionless coupling, the $c_{ijk}$ are the group
structure constants (of $SU(N)$), and the factor of $q^2$ is determined by [2A]
and [2B].

\item[{[2D]}] $t$-channel unitarity is used to determine both $j$-plane
Regge cut discontinuities and particle threshold discontinuities due to
``nonsense'' states.

\item[{[2E]}]  The $j$-plane and $t$-plane discontinuity
formulae are expanded simultaneously around $j=1$ ($\omega = 0$) and in powers
of $g^2$.

\end{description}

The most important ingredient is [2D] - the direct computation of $t$-channel
nonsense state discontinuities. While unitarity determines that reggeon
states produce transverse momentum integrals, we will see that for the
particle discontinuities to be written as transverse momentum integrals it
is necessary that only nonsenses states are involved. When this is the case,
it is straightforward to simultaneously expand the discontinuity equations
in inverse powers of $\omega$ and powers of $g^2$ to determine interactions
etc. in terms of transverse momentum integrals. For example, the $O(g^2)$
contribution to the trajectory function $\Delta(q^2)$ is obtained from the
two-particle discontinuity of the reggeon propagator illustrated in Fig.~2.3

\begin{center}
\leavevmode
\epsfxsize=3in
\epsffile{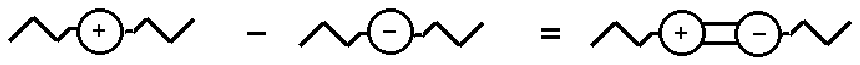}

Fig.~2.3 The two particle discontinuity of the reggeon propagator

\end{center}

\noindent Inserting the structure of the leading-order amplitudes implied by
[2A], [2B] and [2C] we obtain
$$
{1 \over \omega - \Delta(q^2)} ~-~ {1 \over \omega - \Delta^*(q^2)}
{}~=~{ g^2~\sum_{j,k}^N~c_{ijk }c_{ijk}
{}~q^2 \delta_{q^2} \left\{ J_1(q^2) \right\} \over
 (\omega - \Delta(q^2))(\omega - \Delta^*(q^2))}
\auto\label{disc32}
$$
where $\delta_{q^2} \{ ~ \}$ denotes the discontinuity in $q^2$ and
$$
\eqalign{J_1(q^2)~=~{1 \over 16{\pi}^3}
\int {d^2k \over k^2(k-q)^2} }
\auto\label{j1}
$$
{}From (\ref{disc32}) we obtain
$$
\Delta(q^2)~=~g^2~N~q^2J_1(q^2)~+~O(g^4)
\auto\label{g2+}
$$
$O(g^4)$ contributions may come from both the two and
three-particle states.

We obtain the 2-2 reggeon interaction, i.e. the BFKL kernel, from
the nonsense-state discontinuities of the two reggeon propagator Green
function illustrated in Fig.~2.4.

\begin{center}
\leavevmode
\epsfxsize=5in
\epsffile{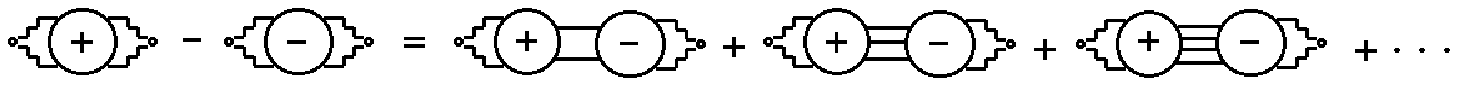}

Fig.~2.4 Nonsense state discontinuities of the two reggeon propagator Green
function
\end{center}
The $O(g^2)$ kernel is derived, as $q^2 \to 0$, from the three-particle
nonsense state. The contributions to the $O(g^4)$ kernel that we derive
will be from the four-particle state. It will be an important part of our
analysis to determine under what kinematic conditions the $t$-channel
discontinuities involved are entirely due to nonsense states.
We will impose not only $q^2 \to 0$
but also limit the individual transverse momenta of the reggeons involved.

\newpage

\mainhead{ 3. THE $O(g^2)$ KERNEL, REGGEON LANGUAGE, WARD IDENTITIES
AND GROUP FACTORS}

Currently the most familiar application of the BFKL equation is to the
evolution of parton distributions at small-x. We begin by recasting this
equation in the reggeon diagram language in which it was originally derived
in order to compare with our results in later Sections.

\subhead{3.1 The BFKL Equation as a Reggeon Bethe-Salpeter Equation}

If $F(x,k^2)$ is a parton distribution then the BFKL equation is
$$
{\partial \over \partial (ln {1 / x})}F(x,k^2) ~=~\tilde{F}(x,k^2)~+~
{1 \over 16\pi^3}\int {d^2k' \over (k')^4} ~K(k,k') F(x,(k')^2)
\auto\label{eve}
$$
where, if $SU(N)$ is the gauge group, $K(k,k')$ is given by
$$
\eqalign{  (Ng^2)^{-1}K(k,q)~=~=~\Biggl(~k^4{k'}^2J_1(k^2)\delta^2(k-k')~
 -~{2k^2{k'}^2 \over (k-k')^2}~\Biggr)}
\auto
$$
To introduce reggeon language we rewrite the equation as a ``reggeon
Bethe-Salpeter equation" by, in effect, working backwards historically. We
first extend (\ref{eve}) to the non-forward direction, then transform to
$\omega$ - space (where $\omega$ is conjugate to $ln~{1 \over x}$), giving
$$
\omega F(\omega,k,q-k) ~=~\tilde{F}~+~ {1 \over 16\pi^3}\int {d^2k' \over
(k')^2(k'-q)^2}~K(k,k',q) F(\omega,k',q-k')
\auto\label{ome}
$$
where $K(k,k',q)~=~K^{(2)}_{2,2}(k,q-k,k',q-k')$ is now the full
``non-forward''
Lipatov kernel and contains three kinematic forms, i.e.
$$
\eqalign{ {2 \over Ng^2}~K^{(2)}_{2,2}(k_1,k_2,k_3,k_4)~
& =~(2\pi)^3k_1^2J_1(k_1^2)k_2^2\Bigl(k_3^2\delta^2(k_2-k_4)
+k_4^2\delta^2(k_2-k_3)\Bigr)\cr
&~~~~~~-~~{k_1^2k_4^2~+~k_2^2k_3^2 \over (k_1-k_3)^2}
{}~~-~~(k_1+k_2)^2\cr
&\equiv~~K_1~+~K_2~+K_3 ~~.}
\auto\label{2,2}
$$

Using a simple notation for transverse momentum diagrams i.e.
using the components illustrated in Fig.~3.1.

\leavevmode
\epsffile{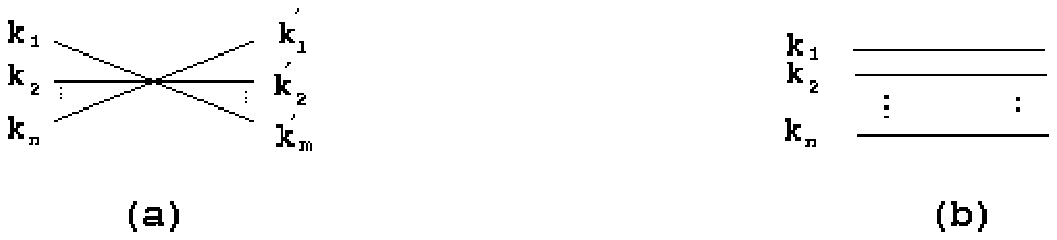}
\begin{center}

Fig.~3.1 (a)vertices and (b) intermediate states in transverse momentum.
\end{center}
the rules for writing amplitudes corresponding to the diagrams are the
following

\begin{itemize}

\item{For each vertex, illustrated in Fig.~3.1(a), we write a factor
$$
16\pi^3\delta^2(\sum k_i~  - \sum k_i')(\sum k_i~)^2
$$
}
\item{For each intermediate state, illustrated in Fig.~3.1(b), we write a
factor
$$
(16\pi^3)^{-n}\int d^2k_1...d^2k_n~ /~k_1^2...k_n^2
$$
}
\end{itemize}
Dimensionless kernels are defined by a hat
$$
\hat{K}^{(2)}_{2,2}(k_1,k_2,k_3,k_4)~=~
16\pi^3\delta^2(k_1+k_2-k_3-k_4) K^{(2)}_{2,2}(k_1,k_2,k_3,k_4)~
$$
The kernels so defined are formally
scale-invariant (even though potentially infra-red divergent). The
diagrammatic representation of $\hat{K}^{(2)}_{2,2}$, the non forward
BFKL kernel, is then as in Fig.~3.2.
\begin{center}
\leavevmode
\epsfxsize=4in
\epsffile{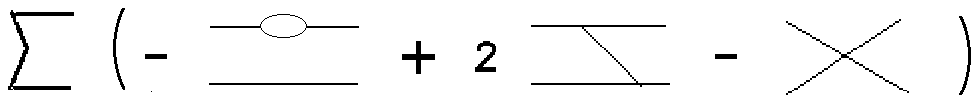}

Fig.~3.2 Diagrammatic representation of $\hat{K}^{(2)}_{2,2}$
\end{center}
The summation sign again implies a sum over combined
permutations of the initial and final momenta.

We introduce a two-reggeon propagator which corresponds to Fig.~2.2 with
$\Delta(k^2)$ given by (\ref{g2+}), i.e. we write
$$
\Gamma_2(\omega,k_1,k_2)~=~[\omega
-g^2k_1^2J_1(k_1^2)-g^2J_1(k_2^2)]^{-1} ,
\auto\label{prop}
$$
moving the $K_1$ term to the left side of (\ref{ome}) and writing
$G~=~{\Gamma_2}^{-1}F$ gives
$$
\eqalign{ G(\omega,k,q-k) ~=~\tilde{G}+
{1 \over (2\pi)^3}\int {d^2k' \over (k')^2(k'-q)^2}~\Gamma_2(\omega,k',q-k')
\tilde{K}(k,k',q) G(\omega,k',q-k') }
\auto\label{bet}
$$
where now
$$
\eqalign{\tilde{K}(k,k',q)~=~ K_2~+K_3~ =~
{k_1^2k_4^2~+~k_2^2k_3^2 \over (k_1-k_3)^2}
-(k_1+k_2)^2 }
\auto\label{int}
$$
can be directly interpreted as a 2-2 reggeon interaction. That this
interaction is singular is what makes it, at first sight, both difficult to
anticipate and to generalize.

Note that if we take $\tilde{G}~=~\tilde{K}(k'',k,q)$ we obtain from
(\ref{bet}) a full reggeon-reggeon scattering amplitude
${\cal G}(\omega,q,k'',k)$ which satifies reggeon unitarity. That is, the
Reggeon propagator in (\ref{bet}) produces a two-reggeon branch-cut in the
$\omega$-plane whose discontinuity is given by
$$
\eqalign{ {\cal G}(\omega^+,q,k'',k)& ~-~{\cal G}(\omega^-,q,k'',k) ~=~
{i \over (2\pi)^2}\int {d^2k' \over (k')^2(k'-q)^2}\cr
&\delta [\omega -g^2k'^2J_1({k'}^2)-g^2(k'-q)^2J_1((k'-q)^2)]
{\cal G}(\omega^+,q,k'',k'){\cal G}(\omega^-,q,k',k) }
\auto\label{betd}
$$
We can represent (\ref{betd}) diagrammatically as in Fig.~3.3.

\begin{center}
\leavevmode
\epsfxsize=4in
\epsffile{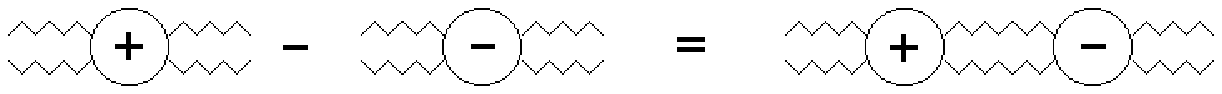}

Fig.~3.3 Reggeon Unitarity for a Reggeon-Reggeon Scattering Amplitude.

\end{center}

\noindent We shall also need the more general discontinuity formula which
follows from (\ref{bet}) for a ``particle-reggeon'' scattering amplitude
$G(\omega,k,q-k)$ defined with a general $\tilde{G}$ i.e.
$$
\eqalign{ G(\omega^+,q,k)& ~-~G(\omega^+,q,k) ~=~
{i \over (2\pi)^2}\int {d^2k' \over (k')^2(k'-q)^2}\cr
&\delta [\omega - g^2k'^2J_1({k'}^2)-g^2(k'-q)^2J_1((k'-q)^2)]
G(\omega^+,q,k'){\cal G}(\omega^-,q,k',k) }
\auto\label{betd1}
$$
This is illustrated in Fig.~3.4.

\begin{center}
\leavevmode
\epsfxsize=4in
\epsffile{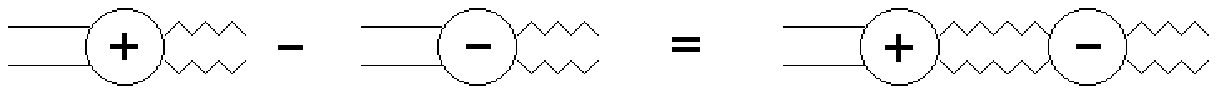}

Fig.~3.4 Reggeon Unitarity for a Particle-Reggeon Scattering Amplitude.

\end{center}
We emphasize that both (\ref{betd}) and (\ref{betd1}) hold
independently of the form of the reggeon interaction $\tilde{K}(k,k',q)$.

If we introduce a Regge slope $\alpha'$ as a regularising
parameter and replace (\ref{prop}) by
$$
\Gamma_2(\omega,k_1,k_2)~=~[\omega -\alpha'k_1^2-\alpha'k_2^2]^{-1}
\auto\label{prop1}
$$
then (\ref{bet}) with $K$ replacing $\tilde{K}$, also reduces to (\ref{ome})
in the limit $\alpha' \to 0$. In this way we can also interpret
(\ref{ome}) directly as a reggeon equation in which the interaction (the full
$K^{(2)}_{2,2}$ ) then has two vital properties

\begin{itemize}

\item It contains singularities (poles) but satisfies the ``Ward Identity
constraint''
$$
\eqalign{ K^{(2)}_{2,2}(k_1,k_2,k_3,k_4) ~~\to ~~0~~, k_i~\to~ 0~, i=~1,..,4}
\auto\label{war}
$$

\item It is infra-red finite as an integral kernel i.e.
$$
\eqalign{ \int {d^2k_1 \over k_1^2} {d^2k_2 \over k_2^2} \delta^2(q-k_1-k_2)
K^{(2)}_{2,2}(k_1,k_2,k_3,k_4) ~~~~~~is ~~finite}
\auto\label{fin}
$$

\end{itemize}

As we pointed out in \cite{ker}, the two properties (\ref{war}) and
(\ref{fin}) determine the relative magnitude of the three kinematic forms
$K_1, K_2,$ and $K_3$ - assuming their existence can be derived from a
general Regge theory argument. We will, of course, provide such an argument in
the following Sections. However, while we will invoke the Ward identity
constraint to determine the relative magnitude of $K_2$ and $K_3$, we will
show how the relative coefficients for $K_1$ and $K_2$ can be determined so
that infra-red finiteness can be derived from unitarity. As we have stated
several times already, and now discuss in more detail, the vanishing of
reggeon amplitudes at zero transverse momentum is our input of gauge
invariance into our analysis.

\subhead{3.2 Gauge Invariance and Reggeon Ward Identities}

As we briefly outlined in \cite{ker}, imposing the vanishing of reggeon
amplitudes at zero tranverse momentum is directly equivalent to imposing the
defining Ward identities of the theory\cite{gth}. We now expand on this
point. In this paper we will always define reggeon amplitudes as residues of
multiple Regge poles in multiparticle partial-wave amplitudes, i.e. we write
$$
a_{j_1,j_2,j_3,j_4,...} ~~\centerunder{$\longrightarrow$}{\raisebox{-3mm}
{$\scriptstyle j_i \to \alpha_i, i=1,..,4$}}~~ \Pi^4_{i=1}~{\beta_i \over
(j_i-\alpha_i)} ~A_{\alpha_1,\alpha_2,\alpha_3,\alpha_1}
\auto\label{rea}
$$
If a Sommerfeld-Watson representation is written\cite{arw1} for the
corresponding multiparticle amplitude the contribution of the reggeon
amplitude to a multi-Regge limit can be obtained. This limit will involve
taking corresponding invariants large, say $s_i \to \infty$ i=1,..,4.
Schematically we can write

\parbox{3in}{\begin{center}
\leavevmode
\epsfxsize=2.5in
\epsffile{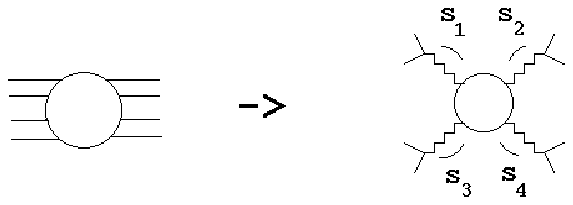}
\end{center}} \ $ \equiv~~\Pi^4_{i=1}
{}~s_i^{\alpha_i}~A_{\alpha_1,\alpha_2,\alpha_3,\alpha_4}$
$$
{}~\auto\label{rea2}
$$
Historically a  reggeon amplitude was generally defined directly this way.

Consider specifically now the reggeon associated with $s_1$. We can always
choose a Lorentz frame in which the limit $s_1 \to \infty$ is defined by
$p_+ \to \infty, k \to k_{\perp}$ where $p$ and $k$ are the momenta labelled
in Fig.~3.5 and $k_{\perp}$ is the transverse momentum carried by the
reggeon.

\begin{center}
\leavevmode
\epsfxsize=4in
\epsffile{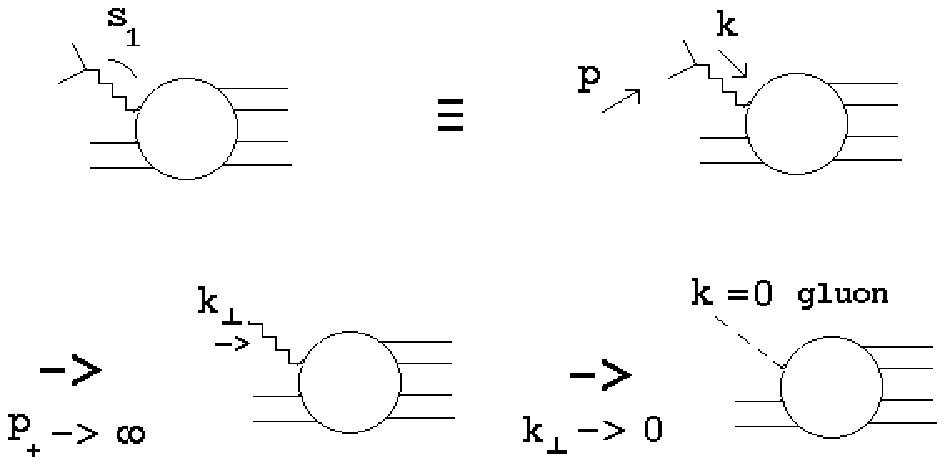}

Fig.~3.5 Reduction of a Reggeon Amplitude to a Gluon Amplitude.

\end{center}

\newpage

\noindent Since the four-momentum $k$ is reduced to a
transverse momentum $k_\perp$ by the Regge limit, the further limit
$k_\perp \to 0$ is equivalent to setting $k=0$. Because of the reggeization
of the gluon, the reggeon amplitude must give the $k=0$ gluon amplitude.
That is, as the transverse momentum of a
reggeon vanishes it can be identified with an elementary gluon carrying {\it
zero four-momentum}. The remainder of the reggeon amplitude under
discussion is embedded in an on-shell S-Matrix amplitude as in (\ref{rea2})
and Fig.~3.5. Therefore we obtain the zero momentum limit of the amplitude
for an off-shell gluon to couple to an S-Matrix element. This amplitude
satisfies a Ward identity\cite{gth}.

The Ward Identity has the form
$$
k_{\mu}~\VEV{A_{\mu}(k)~...~}~=~0
\auto
$$
where $\VEV{A_{\mu}(k)~...~}$ is the amplitude involving a gluon with momentum
$k_{\mu}$. To argue that $\VEV{A_{\mu}~...~}$ vanishes at $k=0$ we simply
differentiate i.e.
$$
\eqalign{ &\VEV{A_{\mu}~...~}~+~{\partial \VEV{A_{\nu}~...~}\over
\partial k_{\mu}}~k_{\nu}~=0\cr
=>~&\VEV{A_{\mu}~...~}~~\centerunder{$\to$} {\raisebox{-5mm}
{$(k_{\mu}\to  0)$}}0~~~~~~~if~~ ~{\partial
\VEV{A_{\nu}~...~}\over \partial k_{\mu}}
{}~\st{\to}~\infty}
\auto\label{wd3}
$$
If there are no
internal infra-red divergences occurring explicitly at zero transverse
momentum (as will be the case in the absence of massless
fermions\cite{arw2}), then, as illustrated, this identity requires the
amplitude to vanish. Clearly this argument can be applied to each of the
reggeons in (\ref{rea2}).

\subhead{3.3 Color Factors}

For our purposes we are interested only in SU(N) gauge theory and so
we can evaluate color factors from the very simple
diagrammatic identities given by C. Y. Lo\cite{cyl}. These are summarized
completely in Fig.~3.6.

\begin{center}
\leavevmode
\epsfxsize=4.5in
\epsffile{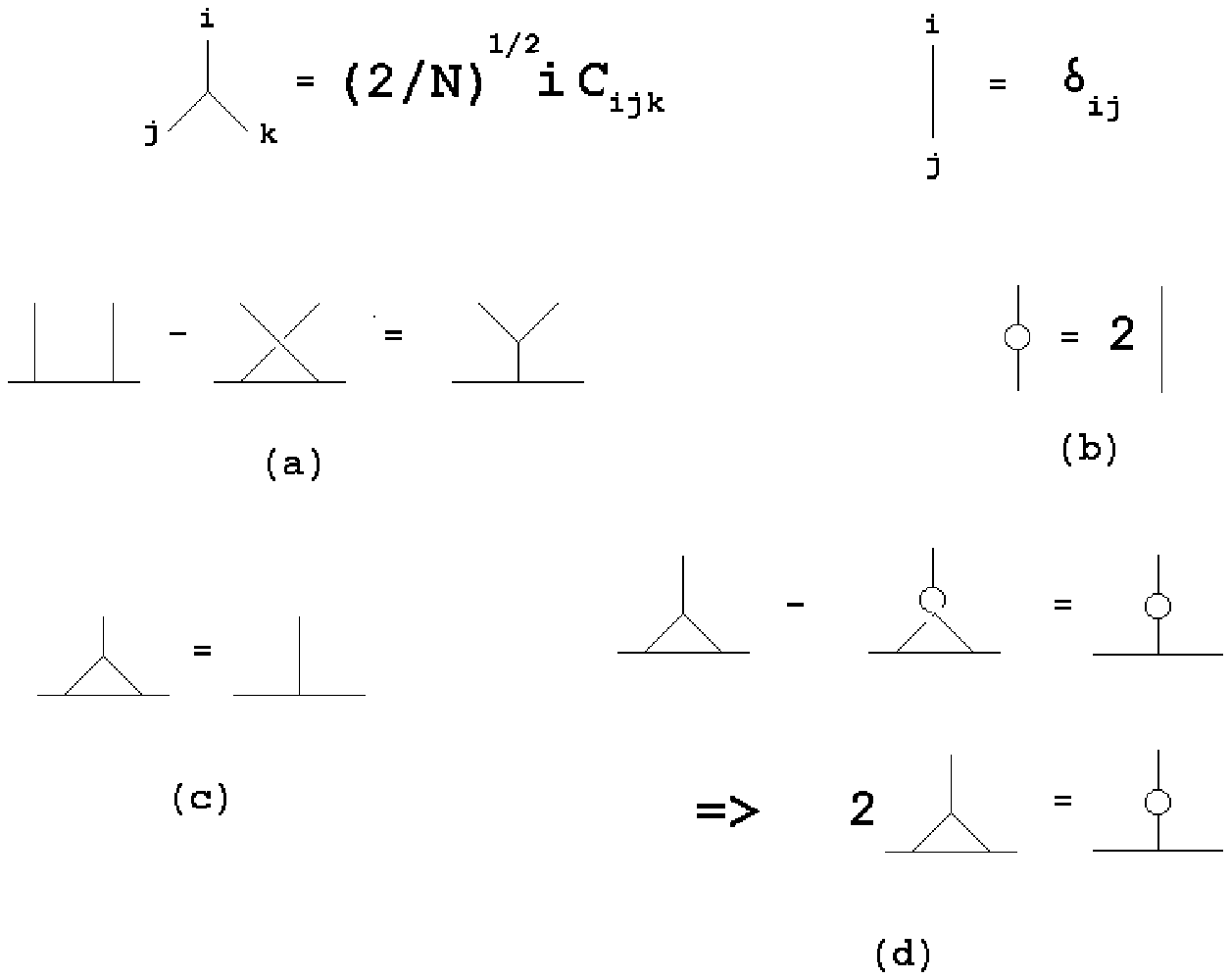}

Fig.~3.6 Structure Constant Identities -  (a) Jacobi Identity (b)
Normalization Relation (c) Triangle Contraction (d) Triangle Contraction
Derivation.
\end{center}

\mainhead{4. NONSENSE REGGEON STATES IN MULTIPARTICLE $t$-CHANNEL
UNITARITY}

In this Section we will try to give a simple, essentially self-contained,
introduction to the elements of Regge theory that are needed in our analysis.
We fear that for those readers with some knowledge of the formalism our
approach will be too elementary, while for those with no knowledge our
description will be far too short on details. Nevertheless we will try to
find the best middle course we can. We will review established results on
Regge cut discontinuities and also discuss reggeization via unitarity.

As we have noted, the analogue of the leading-log expansion in momentum
space is an expansion around $j=1$ in the angular momentum plane. Our aim is
to show how a transverse momentum integral formalism emerges naturally from
analysis of the multiparticle contributions to $t$-channel unitarity near
this point. It will be crucial that, in a vector theory, $j=1$ is a
``nonsense'' point. Consequently we begin by introducing the concept of
``nonsense'' states that occur, by analytic continuation, as ``intermediate
states'' in partial-wave amplitudes. Such states have unphysical (i.e.
nonsense) angular momentum relative to the helicities of the external states
to which they couple. To illustrate this and to introduce the variety of
Regge theory concepts and language we use, we briefly review the formalism
for the simplest possible case, that of elastic scattering.

\subhead{4.1 Wrong-Signature Nonsense Poles, Right-Signature Nonsense Zeroes
and Threshold Behavior}

Consider the partial-wave expansion of an elastic scattering amplitude for
spinless particles i.e.
$$
A(z,t)=\sum^\infty_{j=0}(2j+1)a_j(t)P_j(z),\auto\label{3.1}
$$
$z~=~cos\theta$, where $\theta$ is the $t$-channel center of mass scattering
angle, and the $P_j(z)$ are Legendre polynomials. The inversion formula for
$a_j(t)$ is
$$
\eqalign{
a_j(t)&={1\over 2}\int^{+1}_{-1}dz A(z,t)P_j(z)\cr
&=\int_C dz A(z,t)Q_j(z),}
\auto\label{3.2}
$$
$C$ is a contour enclosing the interval $-1 < z <1$ and so applying
Cauchy's theorem (for $j$ sufficiently large) gives
$$
a_j(t)={1\over 2\pi}\int_{I_R+I_L}dz'Q_j(z')\Delta(z',t)
\auto\label{3.3}
$$
where $I_R$ and $I_L$ are, respectively, the right and left-hand cuts of
$A(z,t)$ and $\Delta(z,t)$ is the corresponding discontinuity. The (unique)
continuation to complex $j$ is given by using
$$
Q_j(z)=(-1)^{j+1}Q_j(-z), ~~~~~~~j=0,1,2,....
\auto\label{3.4}
$$
to define ``signatured'' continuations from even and odd $j$ respectively,
that is
$$
a^\pm_j(t)=\left[a^R_j(t)\mp a^L_j(t)\right]/2,\auto\label{3.5}
$$
where
$$
a^R_j(t)={1\over2\pi}\int_{I^R}dz'Q_j(z')\Delta(z',t)
{}~~~~~~a^L_j(t)={1\over2\pi}\int_{I^L}dz'Q_j(-z')\Delta(z',t).
\auto\label{3.6}
$$
The asymptotic behavior of A(z,t) can be studied via the Sommerfeld-Watson
transformation of (\ref{3.1}) i.e.
$$
A(z,t)~=~\sum_{\pm}~\int dj {(2j+1) \over4 sin\pi j}
a^{\pm}_j(t)\biggl(P_j(z)\pm P_j(-z)\biggr)~.
\auto\label{sw}
$$
where the contour of integration is parallel with the imaginary axis.
Pulling the contour to the left and picking up the singularities of $a_j(t)$
leads to an asymptotic expansion for $z \to \infty$.

Using
$$
Q_{j}(z)~ \centerunder{$\longrightarrow$}{\raisebox{-3mm}
{$\scriptstyle j \to -1$}} ~~\Gamma(j +1) ~\sim~ {1 \over j+1}
\auto\label{3.7}
$$
we have (formally)
$$
a^\pm_j(t) ~\centerunder{$\longrightarrow$}{\raisebox{-3mm}
{$\scriptstyle j \to -1$}}~~ {1 \over 2\pi(j + 1)}~ (\int_{I^R}~ \mp~
\int_{I^L}) dz'\Delta(z',t).
\auto\label{3.8}
$$
Since the external particles have helicity (and spin) zero, $j=-1$ is the first
``nonsense'' point i.e.
$$
j~=~n_1~+~n_2~-~1
$$
where in this case $n_1=n_2=0$. (\ref{3.8}) shows that there is a
``nonsense'' pole which potentially may give a contribution to the
asymptotic behavior.

We now make a crucial observation. If the amplitude is Regge-behaved, that
is the asymptotic behavior is $t$-dependent, then the integrals over
$I^R$ and $I^L$ can be defined for all $t$ values by analytic continuation
from a $t$-range where they converge\cite{arw1}. In this case
$$
(\int_{I^R}~ +~ \int_{I^L}) dz'\Delta(z',t)~=~0
\auto\label{3.9}
$$
- as if the amplitude satisfied an unsubtracted dispersion relation. As a
result the residue of the ``nonsense pole'' given by (\ref{3.8}) vanishes in
the odd (i.e. negative) signature amplitude and so does not contribute to the
asymptotic behavior. In the even-signature case the ``signature factor''
$1+(-1)^j$, arising from $\biggl(P_j(z)\pm P_j(-z)\biggr)$, cancels the
contribution of the pole in the asymptotic expansion obtained from
(\ref{sw}). Noting that $j=-1$ is an odd integer point, we say the
nonsense pole occurs only in the ``wrong-signature'' amplitude (i.e. the even
signature amplitude).

The ``threshold behaviour'' of partial-wave amplitudes will also play an
important role in our discussion. We can illustrate this as follows.
Suppose, for the moment, that the initial and final particles have distinct
massses $m_i, ~i=1,..,4$. In this case we have
$$
\eqalign{s~=~  m_1^2~+&~m_3^2~-~ {(t+m_3^2-m_4^2)(t+m_1^2-m_2^2) \over 2t}\cr
+&{\lambda^{1/2}(t,m_1^2,m_2^2)\lambda^{1/2}(t,m_3^2,m_4^2) \over 2t}~z}
\auto
$$
where, as usual, $\lambda(a,b,c)=a^2 +b^2 +c^2 - 2ab -2bc - 2ac$. Close to
the threshold $\lambda(t,m_1^2,m_2^2)=0$ (or the threshold
$\lambda(t,m_3^2,m_4^2)=0$)  finite $s$ (or $u$) corresponds to large $z$.
Since
$$
Q_j(z)~\centerunder{$\longrightarrow$}{\raisebox{-3mm}
{$\scriptstyle z \to \infty$}}~~ z^{-j-1}
\auto
$$
(\ref{3.3}) then gives
$$
a_j(t)~~~\centerunder{$\sim$}{\raisebox{-5mm}
{$\scriptstyle \lambda(t,m_1^2,m_2^2)~\to~0$}}~
{}~~\biggl({\lambda(t,m_1^2,m_2^2) \over t} \biggr)^{j/2}
{}~\int ds'~{s'}^{-j-1}\Delta(s',t)
\auto\label{thr}
$$

If the external helicities, in the $t$-channel center of mass, are
$(n_1,-n_2)$ for the initial state and $(n_3,-n_4)$ for the final state
then, writing
$n= n_1+n_2$ and $n'=n_3+n_4$, the generalization of (\ref{3.3}) is
$$
\eqalign{~~~~~~~~~a_{jnn'}(t)={1\over 2\pi}\int_{I_R+I_L}dz'
(1+z')^{{n+n' \over 2}}
(1-z')^{{n-n' \over 2}}e^j_{nn'}(z')\Delta(z',t)}
\auto\label{gen}
$$
$e^j_{nn'}(z)$ is a second-type $SO(3)$ representation function
satisfying
$$
\eqalign{~~~~~~~~(1+z')^{{n+n' \over 2}}&(1-z')^{{n-n' \over 2}}e^j_{nn'}\cr
& \centerunder{$\longrightarrow$}{\raisebox{-3mm}
{$\scriptstyle j \to n-1$}} ~
{[\Gamma(j+n+1)\Gamma(j-n+1)\Gamma(j+n'+1)\Gamma(j-n'+1)]^{1/2} \over
2^{1-n}\Gamma(2j+2)}}
\auto\label{gnon}
$$
There is now a ``nonsense branch-point'' at $j=n-1$ rather than
the simple ``nonsense pole'' at $j=-1$. Since there are similar
branch-points in the $d^j_{nn`}(z)$ that replace $P^j(z)$ in (\ref{sw}) the
above discussion of nonsense behavior for spinless amplitudes goes through
analagously, except that (for $n \geq n'$) $j ~\to~ n - 1$ replaces $j \to -1$.
The corresponding generalization of (\ref{thr}) is
$$
a_{jnn'}(t)~~~\centerunder{$\sim$}{\raisebox{-5mm}
{$\scriptstyle \lambda(t,m_1^2,m_3^2)~\to~0$}}~
{}~~\biggl({\lambda(t,m_1^2,m_3^2) \over t} \biggr)^{(j-n)/2}
{}~\int ds'~{s'}^{-j+n-1}\Delta(s',t)
\auto\label{gthr}
$$

For the reggeon amplitudes with non-integer helicities
($n_1=\alpha_1,n_2=\alpha_2$) that appear in the following, the first
nonsense point is $j=\alpha_1+\alpha_2-1$. Both (\ref{gnon}) and (\ref{gthr})
generalize straightforwardly to the case of non-integer helicities. More
details can be found in \cite{arw1}. We move on now to $t$-channel unitarity.

\subhead{4.2 Nonsense Reggeon States and Regge Cuts}

We begin with the application of multiparticle $t$-channel unitarity to
derive Regge cut discontinuities\cite{arw1}. We shall see that nonsense
reggeon states are a key ingredient. Although, as we described in the
Introduction, the history of this formalism goes back thirty years, it is
still not widely known. As an introduction we briefly review the simplest
case, i.e. the derivation of the discontinuity formula for the two-reggeon
cut generated by the exchange of two Regge poles. As we shall see this will
give us directly the general form of the discontinuity formula (\ref{betd})
implied by the BFKL equation.

Consider the the partial-wave projection of the four-particle intermediate
state contribution to the unitarity equation. We use multiparticle
partial-wave amplitudes corresponding to the particular ``coupling scheme''
illustrated in Fig.~4.1.

\begin{center}
\leavevmode
\epsfxsize=2in
\epsffile{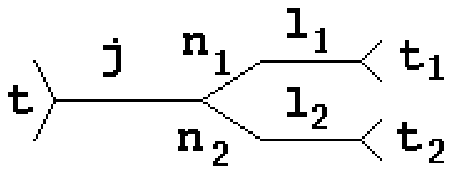}

Fig.~4.1 Partial-wave coupling scheme for the 2-4 production amplitude
\end{center}
$l_1~(l_2)$ and $n_1~(-n_2)$ are respectively the angular momentum
and helicity (in the overall center of mass) of the two-particle state with
energy $t_1~(t_2)$. (Note that we are using the same convention as in
in the previous sub-section. If $n_1$ is a positive helicity in the
center of mass frame, then, $n_2$ is the negative of the corresponding
helicity. With this convention both $j=n_1+n_2-1$ and $j=-n_1-n_2-1$ are
nonsense values of $j$.)

The partial-wave projection of the four-particle unitarity contribution is
$$
\eqalign{~~~~~~~~a_j(t)-a^i_j(t)= \int d\rho \sum_{|n_1+n_2|\leq j}
{}~\sum_{l_1\geq
|n_1|}~ \sum_{l_2\geq |n_2|} a_{j\til{l}\til{n}}(t,\til{t})
a^i_{j\til{l}\til{n}}(t,\til{t})}
\auto\label{proj}
$$
where $i$ denotes an amplitude evaluated on the unphysical side of the
four-particle branch-cut. (We will avoid discussing subtleties associated
with the definition of $i$ amplitudes, in particular the specification of
the additional boundary-values involved.)  If all particles have mass $m$,
but are not identical,
$$
\eqalign{\int d\rho (t,t_1,t_2)~& =~{i \over (2\pi)^52^6} \int dt_1dt_2\cr
&\times \left[{\lambda^{1/2}(t,t_1,t_2)\over
t}\right]\left[{\lambda^{1/2}(t_1,m^2,m^2)\over t_1}\right]
\left[{\lambda^{1/2}(t_2,m^2,m^2)\over t_2}\right]}
\auto\label{pha}
$$
with the integration region defined by $\lambda~\geq~0$, for each of the
three $\lambda$ functions.

The basis for all of the following analysis is the continuation of
unitarity partial-wave projections such as (\ref{proj}) to complex $j$. It
was clear from the original $GPT$ paper\cite{gpt} that this would provide a
very powerful general analysis tool - provided the correct unique form for
the continuation is found. The essential tool we use is (signatured versions
of) the identity
$$
\eqalign{\sum_{{\scriptstyle  n_1 \geq 0,~ n_2 \geq 0
\atop \scriptstyle  j ~\geq~ n_1+n_2 }}~F(j,n_1,n_2)
{}~&=~-~ {sin\pi j \over 2^2} \int_{C_j}
{dn_1dn_2~~F(j,n_1,n_2) \over sin\pi n_1 sin\pi n_2 sin\pi(j- n_1-n_2)}\cr
&\equiv ~-~\Gamma_{[j]}\bigl[ F(j,n_1,n_2)\bigr]}
\auto\label{sin}
$$
which holds when $j$ is an integer, if the integration contour is defined as
a function of $j$ such that, for $j ~\sim~-1/2$, $C_j ~\equiv~
[n_r=-1/4~+i\nu_r~, -\infty < \nu_r < \infty~,~ r=1,2]$, and is defined for
general values of $j$ by analytic continuation. Note that (\ref{sin}) sums
explicitly only the positive values of both $n_1$ and $n_2$. For the full
unitarity equation, each combination of signs has to be treated separately and
gives a distinct contribution to the continuation in $j$. Nonsense states
with $n_1,n_2~>~0$ and $j=n_1+n_2-1$ will appear in the contribution we
discuss explicitly. Nonsense states with $n_1,n_2~<~0$, will appear in a
separate contribution. In much of our discussion we will not refer
explicitly to this and we will implicitly sum over helicity signs.

We will discuss only contributions from odd signature reggeons (gluons) and
so consider only odd-signature values of $n_1$ and $n_2$. In this case we
replace $ sin\pi n_{1,2} $ by $2~sin {\pi \over 2}(n_{1,2} -1)$ in
(\ref{sin}). We will consider both even and odd signature in $j$ and so define
$$
\eqalign{\Gamma^{\pm}_{[j]}~=~{1 \over 2^4}
sin{ \pi \over 2}& (j - \sigma^{\pm})
\int dn_1dn_2 \cr
&\times {1 \over
sin{ \pi \over 2} (j - \sigma^{\pm} -n_1 -n_2 +2)sin{ \pi \over 2}(n_1-1)
sin{ \pi \over 2}(n_2-1)}}
\auto\label{3.20}
$$
where $\sigma^+=0$ and $\sigma^-=1$. $\Gamma^{\pm}_{(j)}$ reduces to a
finite sum when $j$ is an even/odd integer respectively.

To discuss the generation of Regge cuts, we suppose there are Regge poles at
$l_{1,2} = n_{1,2}=\alpha(t_{1,2})\equiv\alpha_{1,2}$ in the production
amplitude partial waves. This is represented diagrammatically in Fig.~4.2

\begin{center}
\leavevmode
\epsfxsize=4.5in
\epsffile{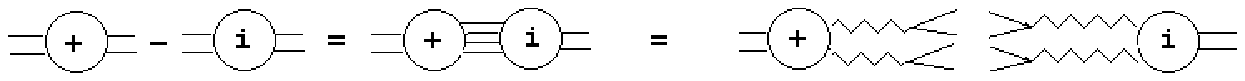}

Fig.~4.2 Regge poles in the production amplitude.
\end{center}
(We should, perhaps, comment that if a Regge pole is present in elastic
amplitudes, its' presence in production amplitudes can actually be
proved\cite{arw1}.) After utilising two-particle unitarity in both the $t_1$
and $t_2$ subchannels, we can write the relevant part of the j-plane
continuation of (\ref{proj}) as
$$
\eqalign{a^{\pm}_j - a_j^{\pm i} =
{}~-~\Gamma^{\pm}_{(j)}\Biggl[~\int d\tilde{\rho}(t,t_1,t_2)
{}~A^{\pm}_{\til{\alpha}}(j,t)A^{\pm i}_{\til{\alpha}}(j,t)
\bigg/ (n_1-\alpha_1)
(n_2-\alpha_2)\Biggr] }
\auto\label{jpl}
$$
where $A^{\pm}_{\til{\alpha}}(j,t)~\equiv~A^{\pm}_{\alpha_1,\alpha_2}(j,t) $
is the production amplitude Regge pole residue and
$$
\int_{C_1,C_2} d\tilde{\rho} (t,t_1,t_2)~ =~{i \over 2^5\pi^3}
\int_{C_1,C_2}  dt_1dt_2\left[{\lambda^{1/2}(t,t_1,t_2)\over
t}\right]
\auto\label{pha1}
$$
As a consequence of the two-particle unitarity manipulations, $C_1$ and
$C_2$ are contour integrals beginning and ending at $\lambda=0$ and
circling, respectively, the thresholds at $t_1=4m^2$ and $t_2=4m^2$. For our
purposes it will be important only that $\lambda=0$ is a boundary of the
integration region. It is important (from a more general point of view)
that the integral is over a finite range of $t_1$ and $t_2$ i.e. it does not
extend to infinite values of the transverse momentum variables that we
introduce below.

We now use the Regge poles at $n_1=\alpha_1$ and $n_2=\alpha_2$ to perform
the $n_1$ and $n_2$ integrations. As we discussed above, there is a
``nonsense branch-point'' factor $(j-\alpha_1-\alpha_2+1)^{-1/2}$ in
$A^{\pm}_{\til{\alpha}}$. For the present discussion, we extract this factor
from both $A^{\pm}_{\til{\alpha}}$ and $A^{\pm i}_{\til{\alpha}}$  and
define (``nonsense'') residue amplitudes $G^{\pm}_{\til{\alpha}}$ and
$G^{\pm i}_{\til{\alpha}}$. We can then rewrite (\ref{jpl}) as
$$
a^{\pm}_j - a^{\pm i}_j ~=~
\Gamma^{\pm}_{(j,t)}\biggl[G^{\pm}_{\til{\alpha}}
G^{\pm i}_{\til{\alpha}}\biggr]  ~~+~...
\auto\label{rct}
$$
where now
$$
\eqalign{\Gamma^{\pm}_{(j,t)}~=&~{{\pi}^2 \over 4} sin{ \pi \over 2}
(j -\sigma^{\pm}) \int { d\tilde{\rho}
\over (j - \alpha_1 -  \alpha_2 + 1)}\cr
&\times~{1 \over sin {\pi \over 2}(j - \alpha_1 -
\alpha_2 +2 -\sigma^{\pm} )sin { \pi \over 2}(\alpha_1 - 1)
sin { \pi \over 2}(\alpha_2 - 1)}}
\auto\label{gam}
$$
We observe that a branch point is potentially generated in
$\Gamma^{\pm}_{(j,t)}$ when the pole at $j = \alpha_1 +
\alpha_2 - 1$ is tangent to the phase space boundary at $\lambda=0$. This
happens when $t_1=t_2=t/4$ and the result is the ``two-reggeon''
branch-point at
$$
j=2\alpha({t \over 4}) -1~.
\auto
$$
(It is important that a branch-point is also generated at $j=2\alpha^*({t
\over 4}) -1$ by that part of the contour below the particle thresholds but
is present in $i$-amplitudes only.)

If the branch-point is indeed to be generated, there must be no
``nonsense zero'' in $G_{\til{\alpha}}$. It follows from the signatured form
of (\ref{gen}) that there is no zero if the signature of $G_{\til{\alpha}}$,
and therefore the signature of the Regge cut, is the product of the
signatures of the participating Regge poles. In particular, if the Regge pole
signatures are identical, the Regge cut has even signature. (Note that other
$j$-dependent ``reggeon-particle'' singularities are also generated in
(\ref{gam}) but only the two-reggeon cut survives\cite{sim,op,arw3} on the
physical sheet for $t \sim 0$).

If neither $G^{+}_{\til{\alpha}}$ or $G^{+i}_{\til{\alpha}}$
contained the Regge cut, the discontinuity would simply be given by
$$
\delta_ja^{+}(j)~=~\delta_j\left\{\Gamma^+_{(j,t)}\right\}~
\biggl[G^{+}_{\til{\alpha}}G^{+i}_{\til{\alpha}}\biggr],
\auto\label{disc}
$$
where $\delta_j$ denotes the discontinuity across the  cut and
$\delta_j\left\{\Gamma^+(j,t)\right\}$ is obtained from $\Gamma^+_{(j,t)}$
by writing
$$(j-\alpha_1-\alpha_2+1)^{-1}~\to~2\pi i
\delta(j-\alpha_1-\alpha_2+1)~.
\auto\label{del}
$$
i.e.
$$
\eqalign{\delta_j\left\{\Gamma^+(j,t)\right\}~=&~{\pi^3 \over 2}
sin{ \pi \over 2}
j \int  d\tilde{\rho} \delta(j - \alpha_1 -  \alpha_2 + 1)\cr
&\times~{1 \over sin {\pi \over 2}(j - \alpha_1 -
\alpha_2 +2 )sin { \pi \over 2}(\alpha_1 - 1)
sin { \pi \over 2}(\alpha_2 - 1)}}
\auto\label{dgam}
$$

In fact $G^{+}_{\til{\alpha}}(j,t)$ does contain the cut, but it is absent
in all $i$-amplitudes. (As we have already remarked, these amplitudes have a
branch-point at the complex conjugate location). Consequently, after a
nontrivial extension of the analysis to amplitudes with reggeons as external
states\cite{arw1}, a standard unitarity manipulation gives the full
discontinuity in the form
$$
\delta_ja^{+}(j)~=~\delta_j\left\{\Gamma^+_{(j,t)}\right\}~
\biggl[G^{+}_{\til{\alpha}}(j^+,t)G^{+}_{\til{\alpha}}(j^-,t)\biggr],
\auto\label{disc1}
$$
where $j^{\pm}$ denotes that the amplitude is evaluated above or below the
reggeon cut involved. The discontinuity formula (\ref{disc1}) holds for
general external states, including the case when they are all reggeons.

To define an amplitude with all external reggeons, i.e. a full reggeon
scattering amplitude, we proceed as anticipated in the last Section. We
first partial-wave project a four-four amplitude according to the coupling
scheme shown in Fig.~4.3.

\begin{center}
\leavevmode
\epsfxsize=3in
\epsffile{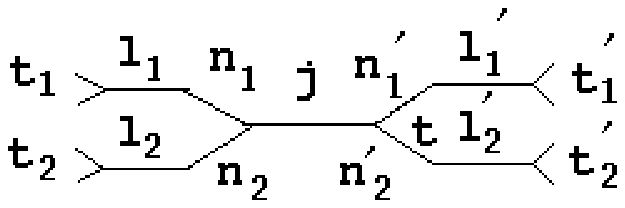}

Fig.~4.3 Coupling scheme for the 4-4 amplitude
\end{center}
After continuation to complex angular momenta and helicities we
then extract the multi-Regge pole residue at $l_1=n_1=\alpha_1,~
l_2=n_2=\alpha_2,~l_1'=n_1'=\alpha_1',~l_2'=n_2'=\alpha_2'$. After
factorising off the couplings of the Regge poles to the external particles
we obtain a reggeon scattering amplitude.

We can write a self-contained ``reggeon unitarity'' equation as follows. We
first define ``nonsense'' reggeon scattering amplitudes by extracting
all the nonsense and threshold factors appearing in (\ref{gnon}) and
(\ref{gthr}). That is we write
$$
\eqalign{ {\cal G}^{\pm}_{\til{\alpha}\til{\alpha'}}(j,t)~=~&
[(j-\alpha_1-\alpha_2+1)(j-\alpha_1'-\alpha_2'+1)]^{1/2}\cr
&\times~\biggl[{\lambda(t,t_1,t_2) \over t}\biggr]^{-(j-\alpha_1-\alpha_2)/2}
\biggl[{\lambda(t,t_1',t_2') \over t}\biggr]^{-(j-\alpha_1'-\alpha_2')/2}
A^{\pm}_{\til{\alpha}\til{\alpha'}}(j,t) }
\auto\label{thr1}
$$
where $A^{\pm}_{\til{\alpha}\til{\alpha'}}(j,t) ~\equiv~
A^{\pm}_{\alpha(t_1),\alpha(t_2),\alpha(t_1'), \alpha(t_2')}(j,t)$ is a full
reggeon scattering amplitude. We then combine the threshold factors with
$\Gamma^+_{(j,t)}$ to define
$$
\tilde{\Gamma}^+_{(j,t)} ~=~\Gamma^+_{(j,t)}
\biggl[{\lambda(t,t_1,t_2) \over t}\biggr]^{(j-\alpha_1-\alpha_2)}
\auto\label{tild}
$$
so that (\ref{disc1}) gives
$$
{\cal G}^+_{\til{\alpha''}\til{\alpha'}}(j^+,t)
{}~-~{\cal G}^+_{\til{\alpha''}\til{\alpha'}}(j^-,t)~=~
\delta_j\left\{\tilde{\Gamma}^+_{(j,t)}\right\}
\biggl[{\cal G}^+_{\til{\alpha''}\til{\alpha}}(j^+,t)
{\cal G}^+ _{\til{\alpha}\til{\alpha'}}(j^-,t)\biggr],
\auto\label{dis2}
$$
where now
$$
\eqalign{\delta_j\left\{\tilde{\Gamma}^+_{(j,t)}\right\}
{}~=&~{\pi \over 2} sin{ \pi \over 2}j
\int  d\tilde{\rho}
\biggl[{\lambda(t,t_1,t_2) \over t}\biggr]^{-1}
{}~\delta(j - \alpha_1 -  \alpha_2 + 1)\cr
&\times~{1 \over sin {\pi \over 2}(j - \alpha_1 -
\alpha_2 +2)sin { \pi \over 2}(\alpha_1 - 1)
sin { \pi \over 2}(\alpha_2 - 1)}}
\auto\label{dtild}
$$
which already compares closely with the discontinuity formula (\ref{betd})
given by the BFKL equation.

We can simplify $\tilde{\Gamma}^+_{(j,t)}$ considerably if we analytically
continue to $t ~\sim 0$. The relevant part of the phase-space is then $t_1
\sim t_2 \sim0$. If we make a linear approximation to the trajectory
function, i.e. $\alpha_r ~= 1 + \alpha't_r + ...$, then we can take
$sin{\pi \over 2}(\alpha_r-1)~\sim~ \pi \alpha't_r/2$ and since $j \sim 1$ we
have $sin{\pi \over 2}j \sim 1$. If we also absorb a factor of
$(\alpha')^{-1}$ in the definition of ${\cal G}_{\til{\alpha''}\til{\alpha'}}$
we obtain by combining (\ref{pha1}) with (\ref{dtild})
$$
\delta_j\left\{\tilde{\Gamma}^+_{(j,t)}\right\}~=~{1 \over 2^4{\pi}^2}
\int  {dt_1dt_2 \over \lambda^{1/2}(t,t_1,t_2)} ~{1 \over
t_1t_2} \delta (j-\alpha_1-\alpha_2 +1)
\auto\label{rct3}
$$
Finally we change to two-dimensional ``transverse momentum'' variables defined,
for our present purposes, by
$$
t=q^2,~~ t_1=k^2,~~t_2=(q-k)^2
\auto\label{trv}
$$
(As we remarked already in the Introduction, the time-like two-dimensional
integrals that we utilise, analytically continue\cite{arw4} into the
corresponding integrals over the usual transverse momentum variables in the
negative $t$ region.)

The jacobian for the transformation (\ref{trv}) is
$$
{dt_1dt_2 \over \lambda^{1/2}(t,t_1,t_2)} ~=~2 d^2k
\auto\label{jac}
$$
If we also write $\omega = j - 1$ and $\Delta_r=1-\alpha_r$ we can write
$$
\delta_j\left\{\tilde{\Gamma}^+_{(j,t)}\right\}~\equiv~
\delta_{\omega}\left\{\tilde{\Gamma}^+_{(\omega,q^2)}
\right\}
\auto\label{var}
$$
where now
$$
\tilde{\Gamma}^+_{(\omega,q^2)} ~=~{1 \over 2^4{\pi}^3}
\int {d^2k \over k^2(k-q)^2} ~{1 \over \omega-\Delta_1-\Delta_2}
\auto\label{rct4}
$$
which we anticipated in (\ref{2rc}), i.e. the discontinuity formula for the
two-reggeon cut is expressed as a transverse momentum integral of the form
that is obtained directly from high-energy calculations. Indeed once
(\ref{var}) and (\ref{rct4}) are utilised, (\ref{betd}) compares directly
with (\ref{dis2}), apart from a factor of i from ${\Gamma}^+_{(j,t)}$ due to
the analytic continuation in t. Note that the threshold factors in
(\ref{tild}), evaluated at the nonsense point, are crucial in producing the
correct jacobian. This explains, from the view-point of $t$-channel
unitarity, why direct Regge-limit calculations produce transverse-momentum
integrals. Such integrals are naturally produced by the exchange of
$t$-channel nonsense states.

As we emphasized in Section 3, the discontinuity formula is satisfied
independently of the form of the kernel. Given that the gluon reggeizes in
leading log, it is then a consequence of $t$-channel unitarity that {\it an
evolution equation, having the general form of the BFKL equation, is
satisfied by the next-to-leading log results.} Our goal is, of course, to
also derive the particular form of the kernel from general unitarity
arguments combined with Ward identity constraints. To do this we will have
to add the extra ingredients of expanding around $j=1$ and extracting
particle threshold singularities in $t$. Neither of which is necessary to
obtain
Regge cut discontinuity formulae.

The above analysis of the two-reggeon cut generalises straightforwardly to
the analysis of the N-reggeon cut - which originates from a nonsense state
of N-reggeons i.e. $j= ~\sum_{r=1}^N\alpha_r~ -N +1$. A self-contained set of
reggeon unitarity equations can be written for multireggeon scattering
amplitudes of both signatures with the signature of the Regge cut due to $N$
reggeized gluons being $(-1)^N$. All the multireggeon discontinuity formulae
can be written in terms of transverse momentum integrals. We emphasize that
this is a property of the phase-space generating the branch-point and is not
a perturbative result. In writing (\ref{rct4}) we implicitly extended the
transverse momentum integrations to infinity. For the discontinuity formula
the upper end point is irrelevant, as it is, analagously, for all
multi-reggeon discontinuity formulae. Since the transverse momentum
integrals that arise in high-energy calculations always extend to infinity we
write our integrals in the same way. Nevertheless it is important to
emphasize that $t$-channel unitarity gives finite integrals. In particular,
when gluons are massless all the structure is at $t=0$

\subhead{4.3 The Trajectory Function as a Nonsense Threshold}

As we now begin to discuss, there are also well-defined reggeon
contributions with signature $(-1)^{N-1}$, i.e. the opposite signature to
the Regge cut they generate. Such contributions give thresholds in $t$
rather than in $j$. Indeed it will be crucial for the emergence of
higher-order kernels from our analysis that reggeons can simultaneously
participate in the generation of Regge cuts and in the generation of thresholds
in $t$. We shall find that in the simplest cases, it is clear that only
nonsense-point kinematics is involved and this leads directly to transverse
momentum integrals for the threshold contributions. However, as we go to
higher-orders this will be a subtle issue that we will need to discuss
carefully. We will set up our general discussion by first describing how the
leading-log reggeization of the gluon can be seen as a threshold contribution
of the two-reggeon state.

{}From (\ref{gam}) we see that there is an additional pole at the nonsense
point in the odd-signature phase-space (from the factor of $sin{\pi \over
2}(j-\alpha_1-\alpha_2+1)$ in the denominator) giving
$$
\eqalign{{\Gamma}^-_{(j,t)}~=~{\pi \over 2}
 sin{\pi\over 2} (j-1) \int { d\tilde{\rho}
\over [j - \alpha_1 -  \alpha_2 +1]^2[sin{\pi\over 2}(\alpha_1-1)]
[sin{\pi \over 2}(\alpha_2-1)]} }
\auto\label{rct5}
$$
In writing (\ref{rct5}) we have extracted nonsense branch-point factors as
in the last sub-section. From (\ref{gnon}) and a generalization of the
dispersion relation argument following (\ref{3.9}), there must be linear
nonsense zeroes in both $G^-_{\til{\alpha}}(j,t)$ and
$G^{-i}_{\til{\alpha}}(j,t)$. These cancel the double pole in (\ref{rct5})
and so no Regge cut is generated. Instead we now focus on the threshold
singularity in $t$ which is generated in ${\Gamma}^-_{(j,t)}$ when
$$
\eqalign{\alpha_1~\equiv~&\alpha(t_1)~=~1~,~~~~
\alpha_2~\equiv~\alpha(t_2)~=~1~,\cr
&\lambda(t,t_1,t_2) ~=~0~.}
\auto\label{tth}
$$
This threshold is similarly generated in the positive signature phase-space
and becomes relevant at non-leading log, although we will not discuss it in
this paper.

(There is a general point underlying the following analysis which we should
note at this point. Two massless gluons can not form a physical massless state
with $j=1$, whereas a nonsense state can have $j=1$ as a result of analytic
continuation. This is why the leading (i.e. most singular) behavior at a
$t$-channel threshold will come from nonsense states coupling via the
three-reggeon vertex - when they can contribute.)

Assuming that the threshold does not appear in $i$-amplitudes
(we discuss conditions under which this will be the case below) we can derive a
discontinuity formula analagous to (\ref{disc}) i.e.
$$
\delta_t\left[{\cal G}^-_{\til{\alpha''}\til{\alpha'}}(j,t)\right]~
=~\delta_t\left\{\tilde{\Gamma}^-_{(j,t)}\right\}~
\biggl[{\cal G}^-_{\til{\alpha''}\til{\alpha}}(j,t^+)
{\cal G}^-_{\til{\alpha}\til{\alpha'}}(j,t^-)\biggr],
\auto\label{dis3}
$$
where ${\cal G}^-_{\til{\alpha''}\til{\alpha}}(j,t)$ is again defined by
(\ref{thr1}) and we have defined $\tilde{\Gamma}^-_{(j,t)}$ in analogy with
the definition of $\tilde{\Gamma}^+_{(j,t)}$ in (\ref{tild}).
We obtain a simple expression for $\tilde{\Gamma}^-_{(j,t)}$ if we set
$\alpha_1=\alpha_2=1$ and consider the leading dependence of (\ref{dis3}) as
$j \to 1$. Since $j=1$ is now the nonsense point relevant for the
phase-space integration, for the leading threshold behavior due to $
\lambda(t,t_1,t_2) \to 0$ we can write, in analogy with
(\ref{rct3})-(\ref{rct4}),
$$
\delta_t\left\{\tilde{\Gamma}^-_{(j,t)}\right\}
{}~\equiv~{1 \over j-1}~\delta_{q^2}\left\{\tilde{\Gamma}^-_{(q^2)}\right\}
{}~=~{1 \over \omega}~\delta_{q^2}\left\{J_1(q^2)\right\}
\auto\label{disc2}
$$
where $J_1(q^2)$ is defined by (\ref{j1}).

Consider now the implications of (\ref{dis3}) when $\alpha_r~\to~1$,  (i.e.
$t_r \to 0$) $~r=1,..,4$, so that $j\sim 1$ is also a nonsense-point for the
external helicities. There are two important features of ${\cal G}^-(j,t)~
\biggl(\equiv{\cal G}^-_{1111}(j,t)\biggr)$ that we have already discussed.
Namely the presence of the reggeized gluon, implying a factor
$(\omega-\Delta(q^2))^{-1}$ and the nonsense zero, implying a factor of
$\omega$. We have not yet discussed the Ward identity constraints.

This is the first point at which it becomes important to impose gauge
invariance on the form of our amplitudes. Because of the Ward identity
constraints discussed in Section 3, the reggeon amplitudes we are
discussing must vanish when any $k_r ~\to 0,~r=1,..,4$. The $k_r$ are two
dimensional momenta with $t_r=k_r^2$. Since we have already set
$k_1^2=k_2^2=k_3^2=k_4^2=0 $, it follows that
$$
q^2~=~(k_1+k_2)^2~=~2k_1.k_2~=~(k_3+k_4)^2~=2k_3.k_4
\auto\label{zer}
$$
and so a factor of $q^2$ will satisfy the constraint. Since ${\cal G}^-$ has
the (transverse momentum) dimensions of $q^2$, we suppose that
$$
{\cal G}^-(j,t)~\equiv~{\cal G}^-(\omega,q^2)
{}~=~{g^2~\omega~q^2 \over (\omega - \Delta(q^2))}
\auto\label{reh}
$$
where $g$ can be treated as a constant. $g$ is defined, at this point, as a
``triple Regge'' coupling of a single reggeized gluon to two gluons in a
nonsense state.

To introduce the ``gauge group'' into our discussion we suppose there is a
global ``color symmetry'' of the reggeon spectrum. In particular we assume
that the reggeized ``gluon'' belongs to the adjoint representation of
$SU(N)$. That is there are $N^2-1$ reggeons with coupling
$c_{ijk}~g,~~i,j,k,~=~1,2,..,N^2-1$. The triple Regge coupling is then
$$
g~c_{ijk}
\auto\label{trvc}
$$
where we identify the $c_{ijk}$ with the usual group structure constants.
(\ref{reh}) becomes
$$
G^-_{i_1,i_2,i_3,i_4}(j,t)~\equiv~G^-(\omega,q^2)
{}~=~{g^2~\sum_{n=1}^N~c_{n~i_1i_2 }c_{n~i_3i_4}~ \omega~q^2 \over (\omega -
\Delta(q^2))}
\auto\label{rehc}
$$
If we reinstate the nonsense factors we have extracted in (\ref{thr1}) (this
will remove the explicit nonsense-zero factor of $\omega$) and set
$\alpha'=0$, this is exactly the nonsense amplitude that is obtained
\cite{gs} by direct $O(g^2)$ calculations.

It is now clear that the lowest-order nonsense reggeon scattering amplitude
(\ref{rehc}) is entirely determined by the following general properties i.e.

\newpage

\begin{itemize}

\item reggeization

\item the color structure of the triple Regge vertex

\item the nonsense zero

\item the Ward identity constraint

\end{itemize}

\noindent each of which is represented by a simple factor. If we
initially take $\Delta(q^2)~=~\alpha'q^2$ then at the Regge pole we can
formally factorize the residue i.e.
$$
g^2~\omega~q^2~\to~g^2~\alpha'(q^2)^2
\auto\label{fac}
$$
and obtain the full triple reggeon vertex
$$
r_{ijk}^0~=~g~c_{ijk}~(\alpha')^{1/2}~q^2
\auto\label{trr}
$$
that we utilised in \cite{ker}. This shows that the momentum factor can be
both interpreted as a nonsense zero and as due to the Ward identity
constraint. We saw in \cite{ker}, and will discuss further shortly, that $g$
can, of course, also be identified as the bare gauge coupling when
comparisons are made with perturbation theory.

As we have defined it, the triple Regge coupling is necessarily
antisymmetric. We have identified it as a 1-2 reggeon coupling, but the
2-reggeon state is not symmetric with respect to interchange of the two
reggeons. It is a nonsense state in which one reggeon carries (center of
mass) helicity +1 and the other carries helicity -1. Interchanging the two
reggeons amounts to a parity transformation - flipping the helicities of all
three reggeons. If the reggeized gluon is a normal vector then this
transformation must produce a change of sign i.e. the coupling should be
antisymmetric. We find it interesting that the antisymmetric nonabelian
gauge coupling is naturally defined as a triple-Regge coupling. It certainly
endorses the concept of defining reggeon non-abelian gauge theories
directly. The inter-relation of the antisymmetry of the coupling with
helicity-flip change of sign will play an important role in the next
Section.

We now look for corrections to the trajectory function of $O(g^2)$. For
simplicity we initially omit the group structure. If we write
$\Delta(q^2)=\alpha' q^2 + g^2\Delta^1(q^2)$ then, from (\ref{disc1}), we
obtain
$$
{1 \over \omega - \Delta(q^2)} ~-~ {1 \over \omega - \Delta^*(q^2)}
{}~=~{ g^2~q^2 \delta_{q^2} \left\{{\Gamma}^-_{q^2}\right\} \over
 (\omega - \Delta(q^2))(\omega - \Delta^*(q^2))}
\auto\label{dtra}
$$
giving directly
$$
\delta_{q^2}\left\{ \Delta^1(q^2) \right\}~
=~q^2~\delta_{q^2}\left\{{\Gamma}^-(q^2)\right\}
\auto\label{disc4}
$$
Clearly the limit $\alpha' \to 0$ is now harmless and the solution of
(\ref{disc4}) is the ``perturbative'' reggeon trajectory. In this simple
case the most obvious solution of the discontinuity formula immediately
gives the full leading-log momentum-space result.
$$
\alpha(q^2)~=~1~+~{g^2 \over 16 {\pi}^3}~q^2~J_1(q^2)
\auto\label{traj}
$$
Including the group structure gives
$$
\eqalign{ \alpha(q^2)~&=~1~+~g^2\sum_{j,k}c^2_{i,j,k}~q^2~ J_1(q^2)\cr
&=~1~+~g^2N ~q^2~J_1(q^2)}
\auto\label{traj1}
$$
where we have used the diagrammatic relation of Fig.~3.5(b). From this last
form of the trajectory function it is immediately clear that $g$ can be
directly identified with the bare gauge coupling.

To prepare for our analysis in the following Sections we note that we could
also have deduced (\ref{disc4}) from a discontinuity formula of the form
$$
\delta_t\left[a(j,t)\right]~
=~\delta_t\left\{\tilde{\Gamma}^-_{(j,t)}\right\}~
\biggl[G_{\til{\alpha}}(j,t^+)
G_{\til{\alpha}}(j,t^-)\biggr],
\auto\label{dis4}
$$
in which $a(j,t)$ is any scattering amplitude containing the gluon Regge
pole. (\ref{disc4}) follows provided only that the coupling of the gluon
pole to the reggeons producing the discontinuity is given by (\ref{trr}).
The pole residues in $a(j,t)$ simply factorize off and are irrelevant,

The original $t$-channel unitarity demonstration of reggeization\cite{gs}
used the two-particle equation continued to nonsense points - with the
particles being those on the reggeon trajectories. The analysis we have just
given is equivalent (apart from the fact that we have determined the
lowest-order nonsense amplitudes from general principles) but generalises
straightforwardly to allow us to discuss ``multiparticle nonsense states''
where the particle lies on a Regge trajectory. As we shall see, it also
allows us to discuss the simultaneous participation of a reggeon in the
generation of a Regge cut and a $t$-channel threshold. Such configurations
will, in higher orders, produce a general structure of threshold
singularities in reggeon interactions.

The reggeon trajectory function can be regarded as the very simplest reggeon
interaction, i.e. the 1-1 interaction. Having extracted the lowest-order
result for this simplest interaction from unitarity we focus on those
features of the analysis that generalize in the following.
The first step was to extract the threshold behavior from the partial-wave
amplitudes and obtain the discontinuity in $t$. Then, with the aid of Ward
identity constraints, we expanded both about the nonsense point $j=1$ and in
powers of $g^2$. (The results, of course, demonstrate that this is
equivalent to the leading-log expansion in momentum space.) The
threshold behavior appearing in (\ref{tild}) implied that we had to be at
the nonsense point to obtain the right jacobian for transformation to
transverse momentum variables. We expect that if we keep higher powers of
$(j-1)$ in the expansion about the nonsense point, we will obtain the
equivalent of momentum space non-leading log amplitudes. Higher nonsense
states will couple with higher powers of $g^2$ and naturally produce
non-leading log interactions. However, from (\ref{tild}) we see that this
expansion will also contain factors of $log[\lambda^{1/2}(t_i,t_j,t_k)]$,
where the $t_{i,j,k}$ are (the square of) momenta carried by reggeons. {\it
To represent such effects in terms of transverse momentum diagrams it will
be necessary to include, as a breaking of scale invariance, logarithms of
the transverse momenta involved.} We will not discuss such effects in this
paper.

So far the role of $\alpha'$ has been only to anticipate
perturbative reggeization effects that ``aposteori'' justify the use
of the reggeon formalism. To recover perturbation theory we
simply set $\alpha'=0$. Justification of our analysis of the $t$-thresholds
actually requires that we both distinguish the particle
thresholds due to reggeons from those appearing in the initial unitarity
equation and have these thresholds appear at distinct locations above and
below the unitarity branch points (below being denoted by ``$i$'' in the
above). To satisfy these requirements we could assume that the unitarity
states we are discussing initially are those of very light particles
(perhaps ``Higgs'' scalars) whose presence in the theory makes the gluons
both massive and unstable so that
$$
\alpha(t)~=~1~- \epsilon ~+~\alpha't~+~...
\auto\label{eps}
$$
If we assume that $\epsilon$ has an imaginary part (due to the light particle
thresholds) the sign of which is reversed below unitarity cuts, then the
above discussion will go through. Our interest is in massless $QCD$ and so
at the end of our analysis we will, of course, assume that both parameters
can be smoothly set to zero. (If we appeal to light particles then this
assumes that they can be decoupled smoothly so that only the unitarity
contributions of gluon reggeons and particles remain. Essentially the
same assumption is made in $s$-channel unitarity calaculations.) As is already
clear from the above, in much of our discussion $\epsilon$ and $\alpha'$
will not play an explicit role and we will omit them. In particular, we will
omit $\epsilon$ entirely.

\mainhead{5. THE BFKL KERNEL FROM THREE REGGEON NONSENSE STATES}

In this Section we recover the $O(g^2)$ kernel described in Section 2 using
the reggeon formalism of the previous Section. We consider the two-two reggeon
interaction produced by the three-reggeon nonsense state. Our procedure will
parallel that used to discuss the trajectory function in the last Section.
We first isolate the three-particle discontinuity formula then, after
constructing the lowest-order nonsense amplitudes from general arguments,
use the discontinuity formula to generate the lowest-order imaginary part of
the interaction.

We consider the six-particle unitarity integral and analyse it with
partial-wave amplitudes corresponding to the coupling scheme shown in Fig.~5.1.

\begin{center}
\leavevmode
\epsfxsize=3in
\epsffile{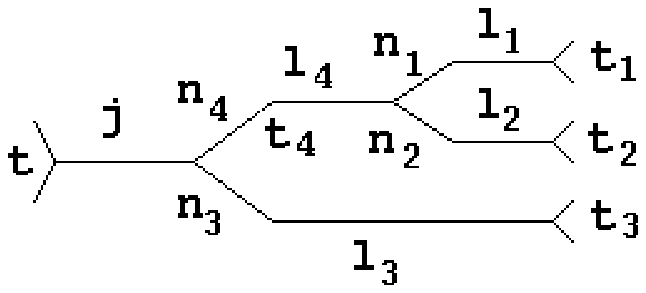}

Fig.~5.1 Coupling scheme for the 2-6 production amplitude
\end{center}
The partial-wave projection of the unitarity integral is
$$
\eqalign{~~~~~~~~a_j(t)-a^i_j(t)= \int d\rho &\sum_{|n_3+n_4|\leq j}
{}~\sum_{|n_1+n_2|\leq l_4}
{}~\sum_{l_1\geq
|n_1|}~ \sum_{l_2\geq |n_2|} ~\sum_{l_3\geq |n_3|}~\sum_{l_4\geq |n_4|}\cr
&\times~
a_{j\til{l}\til{n}}(t,\til{t})
a^i_{j\til{l}\til{n}}(t,\til{t})}
\auto\label{proj6}
$$
where $i$ now denotes amplitudes evaluated below the six-particle cut.
We consider the even signature partial-wave amplitude and keep only the
terms with $l_r~=~n_r,~r=1,..,4$. We will initially pick out the contribution
of
the nonsense pole at $j=n_3+n_4-1$ and the Regge poles at $l_r = \alpha_r
=\alpha(t_r),~ r =1,2,3$, that produce the three-particle threshold.
The two-reggeon cut that appears in $a_j(t)$ is produced in the unitarity
integral by the $\alpha_3$ and $\alpha_4$ Regge poles in combination with
the nonsense pole.

The helicity integrals arising from the continuation to complex $j$ of the
helicity sums in (\ref{proj6}) are (from even signature in $j$ and odd
signature in the $n_r$)
$$
\eqalign{ {1 \over 2^8} sin{ \pi \over 2}j &\int{ dn_3dn_4 \over
sin{\pi \over 2}(j-n_3-n_4)sin{\pi \over 2} (n_3-1)}\cr
&\int { dn_1dn_2 \over
sin{\pi \over 2}(n_4 -n_1 -n_2 +1)sin{ \pi \over 2}(n_1-1)
sin{ \pi \over 2}(n_2-1)}}
\auto\label{6par}
$$
If we again extract the nonsense factors implied by (\ref{gnon}) (we will
discuss shortly why the nonsense points dominate our analysis) we obtain
$$
\eqalign{\Gamma_{3(j)}^+~=~{1 \over 2^7 \pi} &\int{ dn_3dn_4 \over
(j-n_3-n_4+1)
sin{\pi \over 2} (n_3-1)}\cr
&\int { dn_1dn_2 \over
(n_4 -n_1 -n_2 +1)^2 sin{ \pi \over 2}(n_1-1)
sin{ \pi \over 2}(n_2-1)}}
\auto\label{6par1}
$$
Replacing $j$ by $n_4$, the structure of the $n_1$ and $n_2$ integrations is
the same as for $\Gamma^-$ in (\ref{gam}) - except that there is no factor
of $sin{\pi \over 2}(n_4-1)$. Correspondingly, both of the (analytically
continued) partial-wave amplitudes $a_{j\til{l}\til{ n}}$ and
$a^i_{j\til{l}\til{ n}}$ have nonsense zeros at $l_4=n_4=n_1+n_2-1$. These
zeroes eliminate the double pole in (\ref{6par1}) and ensure that no three
reggeon cut is generated. As we discussed in the last Section, and will
exploit below when we extract Regge pole residues, this zero can be
identified with the zero of the triple reggeon vertex.

We now use the Regge poles to perform the helicity integrals over
$n_1,n_2$ and $n_3$ in (\ref{6par1}). We perform the $n_4$ integration by
picking up the nonsense-pole at $j=n_3+n_4-1$. After we use two-particle
unitarity to eliminate the phase-space integrations for the two particle
states to which the $\alpha_1,~\alpha_2$ and $\alpha_3$ Regge poles couple,
the remaining phase-space integration given by the unitarity integral is a
product of integrations of the form of (\ref{pha1}) i.e.
$$
\eqalign{\int d\tilde{\rho} (t,t_1,t_2,t_3,t_4)~=~
{\int}_{\lambda(t,t_3,t_4)>0}~ d\tilde{\rho} (t,t_3,t_4)~
{\int}_{\lambda(t_4,t_1,t_2)>0}~ d\tilde{\rho} (t_4,t_1,t_2)}
\auto\label{pha3}
$$
The only boundaries of the integration region that matter for us are those
we have indicated, at $\lambda(t,t_3,t_4)~=~0$ and $\lambda(t_4,t_1,t_2)~=~0$.

Combining (\ref{6par1}) and (\ref{pha3}) and extracting the nonsense zeroes
we can write the three reggeon contribution to the $j$-plane continuation of
(\ref{proj6}) in the form
$$
\Gamma^+_{3[j,t]}~\tilde{A}_{\til{\alpha}}(j,t)\tilde{A}^i_{\til{\alpha}}(j,t)
\auto\label{uni4}
$$
where $\tilde{A}_{\til{\alpha}}(j,t)$ and $\tilde{A}^i_{\til{\alpha}}(j,t)$
are reggeon amplitudes defined at $n_1=\alpha_1$, $n_2=\alpha_2$,
$n_3=\alpha_3$ and $n_4=j-\alpha_3 +1$ (the notation $\tilde{A}_{...}$ denotes
that the nonsense zero has been extracted) and
$$
\eqalign{\Gamma^+_{3[j,t]}~=~{\pi^3 \over 2^3}
{}~&\int d\tilde{\rho}
{1 \over sin{ \pi \over 2}(\alpha_1-1)
sin{ \pi \over 2}(\alpha_2-1) sin{\pi \over 2} (\alpha_3-1)}}
\auto\label{gam3}
$$

We are interested in the three-particle threshold generated by
$$
\eqalign{\alpha_1~=~\alpha_2~&=~\alpha_3~=~1~,\cr
\lambda(t_4,t_1,t_2) ~&=~\lambda(t,t_3,t_4) ~=~0~.}
\auto\label{tth1}
$$
To isolate the discontinuity associated with the leading behavior we first
extract the threshold factors
$$
{{\cal T}_3}^{1/2}~=~\biggl[{\lambda(t,t_3,t_4) \over t}\biggr]^{(j-n_3-n_4)/2}
{}~\times~
\biggl[{\lambda(t_4,t_1,t_2) \over t_4}\biggr]^{(l_4-n_1-n_2)/2}
\auto\label{thf}
$$
from each amplitude (leaving reduced amplitudes $\tilde{G}$) and absorb them
in the definition of a full phase-space factor
$$
\tilde{\Gamma}^+_{3(j,t)} ~=~\Gamma^+_{3(j,t)}~{\cal T}_3
\auto\label{t3}
$$
When the nonsense conditions $j=n_3+n_4-1$ and $l_4=n_4=n_1+n_2-1$ {\it both
hold,} the threshold factors combine to give the right jacobian factors
to change to transverse momentum variables. In addition to $t=q^2$ we write
$$
\eqalign{ t_1&=k_1^2,~~t_3=k_3^2, ~~t_4=k_4^2=(q-k_3)^2,\cr
{}~~t_2&=k_2^2=(k_4-k_1)^2=(q-k_3-k_1)^2 }
\auto\label{trv1}
$$
In parallel with the discussion
of the previous Section, we use the linear approximation for $\alpha(t)$ and
absorb factors of $\alpha'$ in our definition of residue amplitudes. In
analogy with (\ref{disc2}) we can then write, close to the threshold,
$$
\delta_t\left\{\tilde{\Gamma}^+_{3(j,t)}\right\}
{}~\equiv~\delta_{q^2}\left\{\tilde{\Gamma}^+_{3(q^2)}\right\}
{}~=~\delta_{q^2}\left\{J_2(q^2)\right\}
\auto\label{ttth}
$$
where
$$
\eqalign{J_2(q^2) ~
=~ {1 \over (16\pi^3)^2} ~\int {d^2k_1d^2k_3 \over k^2_1k^2_3(q-k_1-k_3)^2}}
\auto\label{J2}
$$
There is a factor of ${\omega}^{-1}$ missing in (\ref{ttth}) compared to
(\ref{disc2}) because we have already accounted for the presence of nonsense
zeroes. The full (leading-behavior of the) three-particle discontinuity is
$$
\delta_{q^2}\biggl\{a^+(j,q^2)\biggr\}~=~
\delta_{q^2}\bigg\{J_2(q^2) \biggr\}
\tilde{G}_{\til{\alpha}}(j,t^+)\tilde{G}_{\til{\alpha}}(j,t^-)
\auto\label{dis3r}
$$

It is important to discuss why {\it two nonsense conditions hold.} The first
condition, $j=n_3+n_4-1$, holds because we will be considering the two
reggeon nonsense state generating the two-reggeon cut. The second condition,
$l_4=n_4=n_1+n_2-1$, is more subtle. For the leading threshold behavior we
set $n_r = \alpha_r =1,~r=1,..,3$ The second condition then becomes $n_4=1$
while the first condition becomes $j=n_4$. If we then consider $j \sim 1$
the second condition is satisfied. This argument actually demonstrates that
the leading three-particle threshold contribution to the two-reggeon
interaction, i.e. {\it the BFKL kernel, comes entirely from nonsense
states.} For this argument we did not have to impose $q^2 \to 0$ even though
the threshold is at $q^2=0$. Of course, the discussion of the previous
Section also made clear that the simple transverse momentum integrals we
have obtained are only a valid approximation for $j \sim 1$ and $q^2 \sim 0$.

We consider now the lowest order contributions to
$\tilde{G}_{\til{\alpha}}(j,t^+)$ and $\tilde{G}_{\til{\alpha}}(j,t^-)$
(i.e. lowest order in $g$ - the triple reggeon coupling). We consider the
Regge poles at $n_4=\alpha_4$ in $\tilde{G}_{\til{\alpha}}(j,t^+)$ and at
$n_4=\alpha^*_4$ in $\tilde{G}_{\til{\alpha}}(j,t^-)$. As illustrated in
Fig.~5.2,

\begin{center}
\leavevmode
\epsfxsize=4.5in
\epsffile{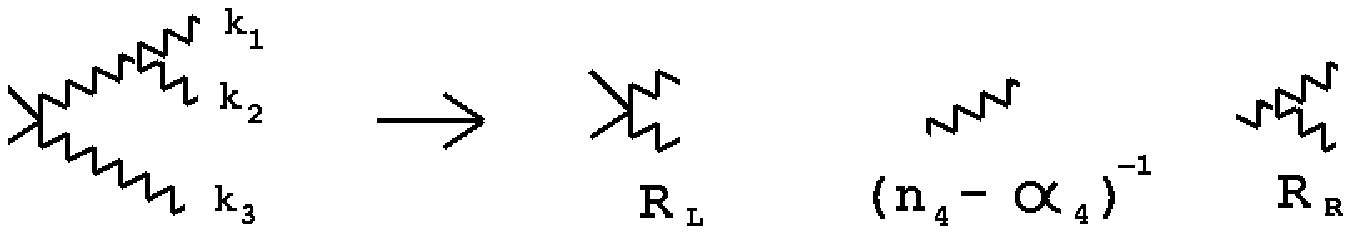}

Fig.~5.2 Factorization of $\tilde{G}_{\til{\alpha}}(j,t^+)$
\end{center}
Regge pole factorization requires that we can write
$$
\tilde{G}_{\til{\alpha}}(j,t^+) ~\sim ~{ R_L~R_R \over n_4 -\alpha_4}
\auto\label{rpf}
$$
where $R_L$ is the coupling of two reggeons to the external state (which we
have taken to be two particles).
For our purposes, we can take $R_L$ to be a constant carrying zero color i.e.
$$
R_L~=~\delta_{ij}
\auto
$$
where $i$ and $j$ are color indices. $R_R$ is the triple reggeon vertex,
except that since we have extracted the nonsense zero there is no momentum
factor. (This momentum factor would satisfy the relevant Ward identity
constraint). Therefore we take
$$
R_R~=~g~c_{ijk}
\auto
$$
Since we have already set $n_4=(j-\alpha_3+1)$ we have
$(n_4-\alpha_4)=(\omega - \Delta_3 - \Delta_4)$ and so we can write
$$
\tilde{G}_{\til{\alpha}}(j,t^+) ~=~{ gc_{ijk} \over
\omega - \Delta_3 - \Delta_4}
\auto\label{rpf1}
$$
(although since we have already set $\alpha_3 = 1$ we could set $\Delta_3 =
0$). With the analagous expression for $\tilde{G}_{\til{\alpha}}(j,t^-)$ we
obtain
$$
\delta_{q^2}\biggl\{ a^+(j,q^2) \biggr\} ~=~
 g^2~C_N~\delta_{q^2}\bigg\{ J_2(q^2) \biggr\}
{1 \over (\omega - \Delta_3 - \Delta_4)(\omega - \Delta^*_3 - \Delta^*_4)}
\auto\label{dis3r1}
$$
$C_N=N$ is the color factor (for SU(N)) obtained by a simple application
of Fig.~3.5(b), as illustrated in Fig.~5.3.

\begin{center}
\leavevmode
\epsfxsize=3.5in
\epsffile{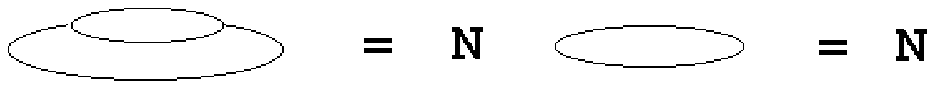}

Fig.~5.3 The color factor $C_N$
\end{center}

Working to $O(g^2)$ in the overall discontinuity, we can neglect the $O(g^2)$
term in $\Delta(q^2)$. We then have
$$
\delta_{q^2}\biggl\{ a^+(\omega,q^2)\biggr\} ~=~
{ g^2~N \over (16\pi^3)^2}
{}~\delta_{q^2} \bigg\{ ~\int {d^2k_1d^2k_3 \over k^2_1k^2_3(q-k_1-k_3)^2}
{1 \over (\omega - \alpha'k_3^2 - \alpha'(q-k_3)^2)^2}
\biggr\}
\auto\label{dis3r2}
$$
This is the discontinuity of the reggeon diagram shown in Fig.~5.4

\begin{center}
\leavevmode
\epsfxsize=2in
\epsffile{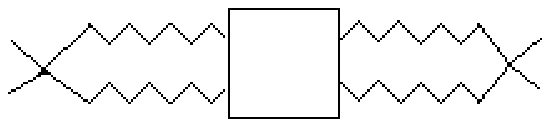}

Fig.~5.4 A reggeon diagram
\end{center}
if we take the reggeon interaction to be
$$
g^2N ~K_1(k_3,q-k_3,k_3',q-k_3')
$$
where $K_1(k_1,k_2,k_3,k_4)$ is given by (\ref{2,2}), corresponding to the
first transverse momentum diagram in Fig.~3.2. Clearly this disconnected
interaction gives again the $O(g^2)$ reggeization that we already
obtained in the previous Section. We have simply rederived this
from the two reggeon state contained within the three reggeon state.

Note that since the two reggeon state includes particle poles (c.f.
(\ref{2rc}) and (\ref{rct3})) the absence of a pole in $k_4^2=(q-k_3)^2$ in
(\ref{dis3r1}) implies that when we rewrite it as a reggeon diagram we
include compensating factors in the interaction. This is equivalent to
reinstating the nonsense zero momentum factor in $R_R$ so that it becomes
the full three reggeon vertex $r^0_{ijk}$ defined in the previous Section.

(\ref{dis3r1}) is not the complete $O(g^2)$ contribution to the
discontinuity. Since we sum over colors for each reggeon we can take
(\ref{rpf1}), in which there is a Regge pole in the $t_4=(k_1+k_2)^2$
channel, to be the complete $O(g)$ contribution to
$\tilde{G}_{\til{\alpha}}(j,t^+)$. However, we must then allow for
contributions to $\tilde{G}_{\til{\alpha}}(j,t^-)$ from Regge poles in the
$t_{23}=(k_2+k_3)^2$ and $t_{13}=(k_1+k_3)^2$ channels. The reggeon amplitude
with a pole in $t_{23}$ is shown in Fig.~5.5.

\begin{center}
\leavevmode
\epsfxsize=1.3in
\epsffile{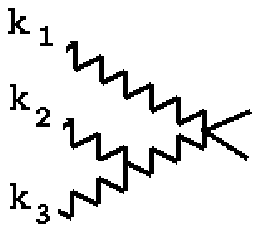}

Fig.~5.5 A reggeon amplitude
\end{center}
In principle we should directly evaluate the contribution of
this amplitude to the partial-waves that we have used up to this point.
In general this would be a complicated transformation. However, as we now
discuss, there are special kinematic situations in which the transformation
simplifies.

The amplitude of Fig.~5.5 has a simple form in the partial-wave coupling
scheme illustrated in Fig.~5.6.
Let us compare, in general, the variables corresponding to
Fig.~5.1 and Fig.~5.6.
\begin{center}
\leavevmode
\epsfxsize=3in
\epsffile{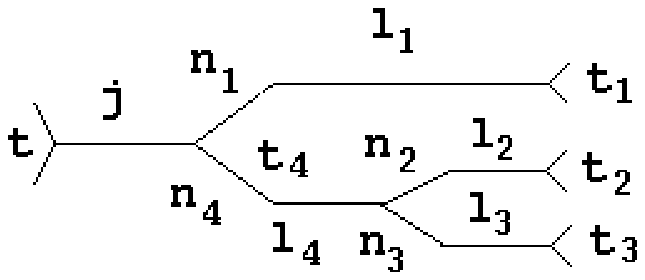}

Fig.~5.6 Alternative coupling scheme
\end{center}
For Fig.~5.1 the angular variables conjugate to the
angular momenta and helicities are defined\cite{arw1} with respect to the
plane in which the momenta of $t$ and $t_1$ lie and the plane given by $t_2$
and $t_3$. The variables of Fig.~5.6 are defined with respect to the plane of
$t$ and $t_3$ and that of $t_2$ and $t_3$.

Consider now the leading threshold behavior at $t= q^2 = 0$. To obtain
$q^2=0$ from three ``massless'' particles, i.e. with $k_i^2 = 0$, $i=$
1,2,3, {\it all three momenta must be parallel.} This implies that {\it in this
special case} the relevant variables of Figs.~5.1 and 5.6 degenerate. The
helicities of the three particles can be identified, the angles conjugate to
$j$ and $n_4$ can essentially be identified within each scheme and also in
the two schemes. All singularities of
the partial-wave amplitudes are associated with the threshold factors that
we have already extracted. Consequently the reggeon
amplitude of Fig.~5.6 can be simply expressed in terms of the variables of
Fig.~5.1. That is we can write the contribution of the diagram of Fig.~5.5
to $\tilde{G}_{\til{\alpha}}(j,t^-)$ in the form
$$
\tilde{G}_{\til{\alpha}}(j,t^-) ~\sim~ { R_L~R_R \over \omega - \Delta^*_1 -
\Delta^*_{23}}
\auto\label{rpf2}
$$
where $\Delta_{23} = \alpha'(k_2 + k_3)^2$. Now we take $R_R~=~\delta_{ij}$
and $R_L$ is the triple reggeon vertex. We have only to determine the
relative sign, which we do by the following argument based on the
antisymmetry properties of the two-reggeon state discussed earlier.

We need the relative contribution of the two reggeon diagrams shown in
Fig.~5.7.
{}From the analysis of the two-reggeon cut in the previous Section,
we know that the reggeons forming the two reggeon state
coupling to the particles must have opposite helicities. Since helicity is
conserved by the reggeons in a reggeon diagram, the triple reggeon vertices
in the two diagrams involve single reggeons of opposite helicity. That is
one vertex is the parity transformation of the other and so must have the
opposite sign (when all color labels are identical).

\begin{center}
\leavevmode
\epsfxsize=2.5in
\epsffile{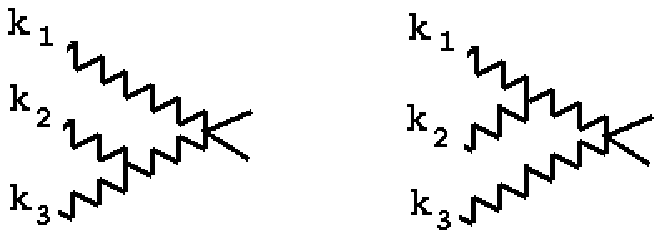}

Fig.~5.7 Comparison of reggeon diagrams
\end{center}
We therefore take
$$
R_L~=~-c_{ijk} \auto
$$
Inserting (\ref{rpf1}) and (\ref{rpf2}) in
(\ref{dis3r}) we again obtain the discontinuity of a reggeon diagram of the
form shown in Fig.~5.4 if we take the reggeon interaction to be
$$
g^2N ~K_2(k_3,q-k_3,k_3',q-k_3')
$$
with $K_2(k_1,k_2,k_3,k_4)$ given by (\ref{2,2}) and corresponding to the
second diagram of Fig.~3.2 - apart from a factor of
two which is obtained by adding the diagram with the pole in the
$t_{12}$-channel. The color factor is obtained via the appplication of
Fig.~3.5 illustrated in Fig.~5.8.
\begin{center}
\leavevmode
\epsfxsize=4in
\epsffile{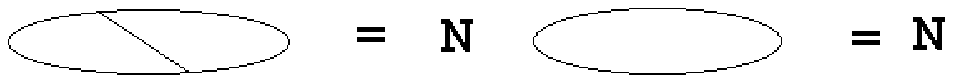}

Fig.~5.8 Another color factor
\end{center}

Since we have determined both the overall magnitude and relative sign of
$K_1$ and $K_2$ (the overall sign actually changes as we continue from
timelike to spacelike $q^2$), the infra-red finiteness property of Section 3
has been obtained directly from unitarity. $K_1 + K_2$ is the complete
kernel at $q^2=0$. $K_3$ vanishes at this point and since it has no
discontinuity in $q^2$ can not be determined by unitarity. It is immediately
determined as the first correction away from $q^2=0$ once we impose the Ward
identity constraint (\ref{war}) that is our input of gauge invariance.
Therefore the full, conformally invariant , BFKL kernel is determined by the
combination of $t$-channel unitarity and Ward identity constraints.

\mainhead{6. FOUR-REGGEON NONSENSE STATES}

We now have all the apparatus in place to discuss the derivation, from the
four-reggeon nonsense states, of the components of $K^{(4n)}_{2,2}$, the
$O(g^4)$ kernel discussed in \cite{cw}. We recall that $K^{(4n)}_{2,2}$ was
defined by the sum of transverse momentum integrals
$$
\eqalign{{1 \over (g^2N)^2} K^{(4n)}_{2,2}(k_1&,k_2,k_3,k_4)
{}~=~K^{(4)}_0~+~K^{(4)}_1~+~K^{(4)}_2~+~K^{(4)}_3~+K^{(4)}_4~}.
\auto\label{sum}
$$
with
$$
\eqalign{K^{(4)}_0~=~
\sum ~ k_1^4k_2^4J_1(k_1^2)J_1(k_2^2)(16\pi^3)\delta^2(k_2-k_3)~,}
\auto
$$

$$
\eqalign{K^{(4)}_1~=~-{2 \over 3}~
\sum ~ k_1^4J_2(k_1^2)k_2^2(16\pi^3)\delta^2(k_2-k_3)}
\auto
$$

$$
\eqalign{K^{(4)}_2~=~- ~\sum
\Biggl({k_1^2J_1(k_1^2)k_2^2k_3^2+
k_1^2k_3^2J_1(k_4^2)k_4^2 \over
(k_1-k_4)^2} \Biggr),}
\auto
$$

$$
\eqalign{K^{(4)}_3~=~\sum~
k_2^2k_4^2J_1((k_1-k_4)^2)~,}
\auto
$$
and
$$
\eqalign{K^{(4)}_4~=~{1 \over 2}~\sum~
k_1^2k_2^2k_3^2k_4^2~I(k_1,k_2,k_3,k_4), }
\auto
$$
The corresponding transverse momentum diagrams are shown in Fig.~6.1

\begin{center}
\leavevmode
\epsfxsize=5in
\epsffile{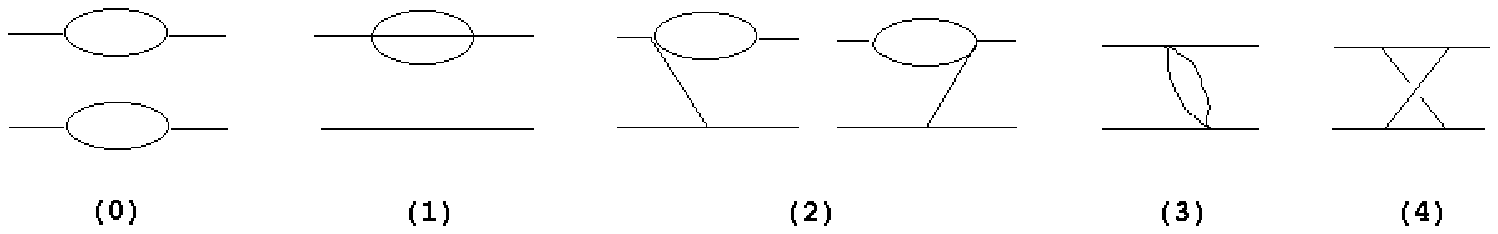}

Fig.~6.1 (0), (1) - disconnected diagrams for the $O(g^4)$
kernel; (2), (3), (4) - connected diagrams.

\end{center}

We will encounter essentially all of the subtleties of multiparticle
multi-Regge theory in this Section and although we will try to give a
coherent self-contained discussion it is likely that \cite{arw1} is an
essential reference to follow the full details. We use the partial-wave
coupling scheme shown in Fig.~6.2.

\begin{center}
\leavevmode
\epsfxsize=3in
\epsffile{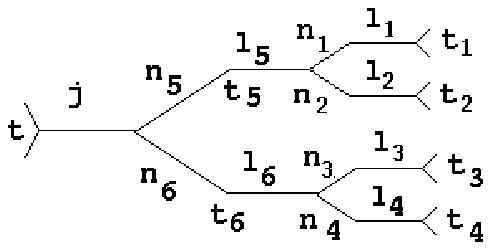}

Fig.~6.2 Coupling scheme for the eight-particle state
\end{center}
The partial-wave projection of the unitarity integral is
$$
\eqalign{~~~~~a_j(t)-a^i_j(t)= \int d\rho &\sum_{|n_5+n_6|\leq j}
{}~\sum_{|n_1+n_2|\leq l_5} ~\sum_{|n_3+n_4|\leq l_6}
{}~\sum_{r=1}^6 ~\sum_{l_r\geq |n_r|}\cr
&\times~
a_{j\til{l}\til{n}}(t,\til{t})
a^i_{j\til{l}\til{n}}(t,\til{t})}
\auto\label{proj8}
$$
where now the $i$ denotes amplitudes evaluated below the eight-particle cut.

As before, we consider even signature in $j$,  and keep the terms with
$l_r~=~n_r,~r=1,..,6$. The helicity integrals giving the continuation to
complex $j$ of the sums in (\ref{proj8}) are (keeping only odd signature in
the $n_r$)
$$
\eqalign{ &{1 \over 2^{12}} sin{ \pi \over 2}j \int{ dn_5dn_6 \over
sin{\pi \over 2}(j-n_5-n_6)}\cr
&\int { dn_1dn_2 \over
sin{\pi \over 2}(n_5 -n_1 -n_2 +1)sin{ \pi \over 2}(n_1-1)
sin{ \pi \over 2}(n_2-1)}\cr
&\int { dn_1dn_2 \over
sin{\pi \over 2}(n_6 -n_3 -n_4 +1)sin{ \pi \over 2}(n_3-1)
sin{ \pi \over 2}(n_4-1)}
}
\auto\label{8par}
$$
This time the structure of both the $n_1$ and $n_2$ integrations and the
$n_3$ and $n_4$ integrations is the same as for $\Gamma^-$ in (\ref{gam}).
Correspondingly nonsense zeroes (at the nonsense points we discuss below)
ensure that no internal two-reggeon cuts are generated that would lead to a
four-reggeon cut. As we discussed in the last Section, this zero can be
identified with the zero of the triple reggeon vertex. The threshold we
consider is produced by the Regge poles at $l_r = \alpha_r =\alpha(t_r),~ r
=1,..,4$. The two-reggeon cut is produced by Regge poles at $l_5 = \alpha_5$
and $l_6 = \alpha_6$, together with the nonsense pole at $j=n_5+n_6-1$.

We use the Regge poles to perform the helicity integrals over
$n_1,n_2$, $n_3$ and $n_4$ in (\ref{8par}) and, as usual, use two-particle
unitarity to eliminate the phase-space integrations for the two particle
states to which the $\alpha_1,\alpha_2,\alpha_3$ and $\alpha_4$ Regge poles
couple. Retaining for the moment the integrations over $n_5$ and $n_6$ and
extracting nonsense factors we obtain, for $j \sim 1$,
$$
\eqalign{~~~~~\Gamma^+_{4[j,t]}~\sim &{\pi^4 \over 2^6} \int {dn_5dn_6 \over
(j-n_5-n_6+1)}\cr
&\int d\tilde{\rho}
{1 \over sin{ \pi \over 2}(\alpha_1-1) sin{ \pi \over 2}(\alpha_2-1)
sin{\pi \over 2} (\alpha_3-1)sin{ \pi \over 2}(\alpha_4-1) }}
\auto\label{gam4}
$$
where $\int d\tilde{\rho}$ is a product of integrations of the form of
(\ref{pha1}) i.e.
$$
\eqalign{~~~\int d\tilde{\rho} (t,t_1,...,t_6)~=~&
{\int}_{\lambda(t,t_5,t_6)>0}  d\tilde{\rho} (t,t_5,t_6)\cr
&{\int}_{\lambda(t_5,t_1,t_2)>0}~ d\tilde{\rho} (t_5,t_1,t_2)
{\int}_{\lambda(t_6,t_3,t_4)>0}~ d\tilde{\rho} (t_6,t_3,t_4)}
\auto\label{pha4}
$$
The boundaries of the integration region that matter for us are those
we have shown. $\lambda(t,t_5,t_6) = 0$~ is involved in both the two-reggeon
cut and the four-particle threshold. $\lambda(t_5,t_1,t_2) = 0$ and
$\lambda(t_6,t_3,t_4) = 0$ will contribute to the four-particle threshold.

We are, of course, interested in studying the four-particle threshold in
combination with the two-reggeon cut. If the phase-space (\ref{pha4}) is to
reduce to transverse momentum integrals then, in addition to the nonsense
condition imposed by the two-reggeon cut, nonsense conditions must be
satisfied at the two ``internal vertices'' i.e. we must have
$$
l_5~=~n_5~ \sim ~n_1+n_2-1~~, ~~~~~~~~~~~~~~l_6~=~n_6~ \sim ~n_3+n_4-1~~,
\auto\label{nons}
$$
Extracting the leading threshold behavior will set $n_1=n_2=n_3=n_4=1$ and so
(\ref{nons}) will be satisfied if $n_5\sim n_6 \sim1$. If we consider
$t \sim 0$ then the two-reggeon cut is generated at $t_5 \sim t_6 \sim t/4 $
which, provided the reggeon slope $\alpha'$ is finite, implies
also that $n_5 \sim n_6 \sim~1$, as we
require. Therefore, if $\alpha'$ is finite and we are interested strictly in
the two-reggeon cut, $j ~\sim ~ 1$ is equivalent to
$$
t ~\sim ~t_5~ \sim ~t_6 ~\sim~ (j-1)/ \alpha'~\sim~ 0 ~.
\auto\label{kin}
$$
It follows that, for $\alpha'~\neq ~0$, the leading four-particle threshold
contribution to the two-reggeon interaction, in the neighborhood of $j~=~1$,
is indeed given by transverse momentum integrals. However, we are interested
in the limit $\alpha' \to 0$, which in principle allows $t, t_5, t_6$ to be
arbitrarily large while satisfying (\ref{kin}). Nevertheless we shall find that
there are additional kinematic arguments that lead us to impose
$t_5 \sim t_6 \sim 0$, in addition to  $t \sim 0$ and $j\sim 1$. In this case
the nonsense conditions remain valid as $\alpha' \to 0$. Clearly we will
obtain only an infra-red approximation to the reggeon interaction we are
looking for.

Extracting threshold behavior ${\cal T}_4$ in analogy with (\ref{thf})
and defining a full phase-space factor
$$
\tilde{\Gamma}^+_{4(j,t)} ~=~\Gamma^+_{4(j,t)}~{\cal T}_4
\auto\label{t4}
$$
we can write, in parallel with the discussion of previous Sections, except
that we still retain the integrations over $n_5$ and $n_6$,
$$
\eqalign{ \delta_t\left\{\tilde{\Gamma}^+_{4(j,t)}\right\}
{}~&\equiv~\delta_{q^2}\left\{\tilde{\Gamma}^+_{4(q^2)}\right\}\cr
&=~\delta_{q^2}\left\{J_3(q^2)\Biggl[{\pi^4 \over 2^6} \int {dn_5dn_6 \over
(j-n_5-n_6+1)}\Biggr]\right\}}
\auto\label{tfth}
$$
where
$$
\eqalign{J_3(q^2)\Biggl[...\Biggr] ~
=~ {1 \over (16\pi^3)^3} ~\int {d^2k_1d^2k_3d^2k_4 \over k^2_1k^2_3k^2_4
(q-k_1-k_3-k_4)^2}~~\Biggl[...\Biggr]}
\auto\label{J3}
$$
The transverse momenta are now defined by
$$
\eqalign{ t_i~&=~k_i^2 ~~~~~~~~~~~i=1,...,6\cr
k_2~&=~q-k_1-k_3-k_4\cr
k_5~&=~k_1+k_2\cr
k_6~&=~k_3+k_4 }
\auto\label{trv4}
$$
The leading-behavior of the four-particle discontinuity is then, formally,
$$
\delta_{q^2}\biggl\{a^+(j,q^2)\biggr\}~=~
\delta_{q^2}\bigg\{J_3(q^2)\Biggl[
{\pi^4 \over 2^6} \int {dn_5dn_6 ~
\tilde{G}_{\til{\alpha}}(j,t^+)\tilde{G}_{\til{\alpha}}(j,t^-)
\over (j-n_5-n_6+1)} ~~\Biggr] ~\biggr\}
\auto\label{dis4r}
$$
where $\tilde{G}_{\til{\alpha}}(j,t)$ is now a two-particle/four-reggeon
amplitude.

At first sight $K^{(4)}_0$ is the simplest to derive. It is certainly the
simplest to discuss. Since it is a sum of diagrams of the form of
Fig.~6.1(0), it should be generated by a diagonal product of partial-wave
amplitudes as a straightforward generalization of the discussion of $K_1$ in
the BFKL kernel. Instead of Fig.~5.2, we have the factorization illustrated
in Fig.~6.3

\begin{center}
\leavevmode
\epsfxsize=4in
\epsffile{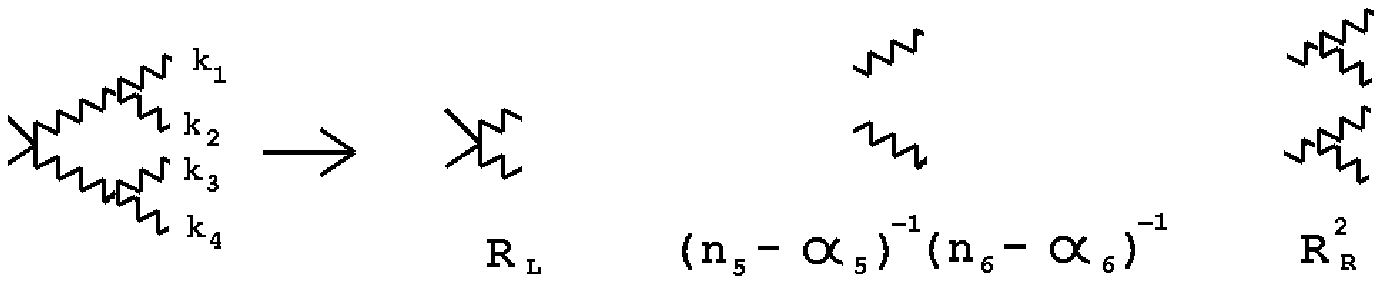}

Fig.~6.3 Factorization of $\tilde{G}_{\til{\alpha}}(j,t^+)$
\end{center}
that is we write
$$
\tilde{G}_{\til{\alpha}}(j,t^+) ~\sim ~{ R_L~R_R^2 \over (n_5 -\alpha_5)
(n_6 -\alpha_6)}
\auto\label{rpf3}
$$
where $R_R$ is the triple Regge vertex. The contribution from
$\tilde{G}_{\til{\alpha}}(j,t^-)$ is analagous.

However, the poles at $n_5 = \alpha_5$ and $n_6 = \alpha_6$
together with the complex conjugate poles in $\tilde{G}_{\til{\alpha}}(j,t^-)$,
give for the helicity integral in  (\ref{dis4r})
$$
\int { dn_5dn_6 \over
(j-n_5-n_6 +1)(n_5 -\alpha_5)(n_6 -\alpha_6)
(n_5 -\alpha_5^*)(n_6 -\alpha_6^*)}
\auto\label{k0p}
$$
Using $(j-n_5-n_6 +1)^{-1}$ for one integration, gives
$$
\eqalign{ {1 \over (\alpha_6-\alpha_6*)} \Biggl(& {1 \over
(j-\alpha_5 -\alpha_6 +1)(j-\alpha_5^* -\alpha_6 +1)}\cr
 & -~{1 \over
(j-\alpha_5 -\alpha_6^* +1)(j-\alpha_5^* -\alpha_6^* +1)} \Biggr)}
\auto\label{k0p1}
$$
Since two reggeon propagators appear in each term we could, after insertion
back in (\ref{dis4r}), extract the residue of one term as (the discontinuity
of) a reggeon interaction. (We would get the same result
from either term since we would make the approximation $\alpha_5 = \alpha_5^*$
and identify the first term with $a_j(t^+)$ and the second with $a_j(t^-)$).
However, we do not obtain $K^{(4)}_0$. The additional factor of
$(\alpha_6-\alpha_6*)^{-1}$ in (\ref{k0p1}) has the effect of removing the
corresponding $J_1$ factor in Fig.~6.1(0). Consequently we do not produce
$K^{(4)}_0$, but again reproduce the lowest-order reggeization $K_1$. (There
is not actually a pole at $\alpha_6 = \alpha_6*$ since the two terms in
(\ref{k0p1}) cancel at this point.)

It should be no surprise that $K^{(4)}_0$ is not generated via a
scale-invariant unitarity analysis. It has no interpretation as a trajectory
function contribution and as a reggeon diagram it is completely dependent on
the existence of a rapidity-gap minimum cut-off which defines the ``finite
rapidity interval'' within which the two bubble interactions of $K^{(4)}_0$
can occur. If this cut-off is set to zero, as effectively is done in the
above unitarity analysis, the diagram disappears.

The situation is different for the next contribution we discuss. We consider
$K^{(4)}_4$, corresponding to the diagram of Fig.~6.1(4). This is an
off-diagonal product of the same reggeon diagrams that at first sight produce
$K^{(4)}_0$. To discuss this we need, in principle, to obtain an expression for
the reggeon diagram of Fig.~6.4 in the partial-wave coupling scheme of
Fig.~6.2.
\begin{center}
\leavevmode
\epsfxsize=1.3in
\epsffile{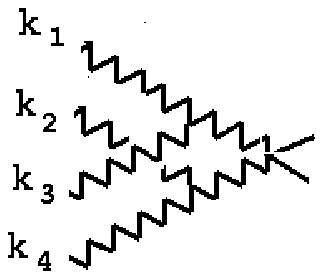}

Fig.~6.4 A reggeon contribution to $\tilde{G}_{\til{\alpha}}(j,t^-)$.
\end{center}
As in our discussion of Fig.~5.5 in the last Section, a complicated
transformation is involved in general. We can again impose special kinematic
restrictions to simplify the situation.

If we impose {\it both} $t = (k_1 + k_2 + k_3 + k_4)^2 =0$ and, say,
$(k_1 + k_3)^2 = 0$ on four massless particles (allowing the $k_i$ to be
four momenta) then, necessarily, {\it all the momenta are parallel}. Therefore
the angular variables of Fig.~6.2 and the coupling scheme associated with
Fig.~6.4 degenerate. (All the $\lambda$-functions in (\ref{pha4}) vanish and
the divergences of the partial-wave amplitudes are
again extracted by the threshold factors involved in changing to transverse
momenta.) However, there is a further problem not present in our discussion
of Fig.~5.5. In the reggeon diagram of Fig.~6.4, the variables $(k_1 +
k_3)^2$ and $(k_2 + k_4)^2$ are clearly transverse momenta and also are zero
when all the momenta are parallel. In terms of the variables
associated with the coupling scheme of Fig.~6.2, these ``transverse momentum
variables'' are expressed in terms of angular variables that are the
analogue of rapidity variables in the $s$-channel. Consequently taking $(k_1
+ k_3)^2 \sim 0$ and $(k_2 + k_4)^2 \sim 0$ limits the angular range (the
relative rapidity of the $(t_5,t_1,t_2)$ and $(t_6,t_3,t_4)$ vertices)
integrated over when projecting Fig.~6.4 onto the partial-waves of Fig.~6.2.
This range determines the normalization of the contribution of Fig.~6.4.
Thus defining the transverse momentum scale implied by writing
$$
t = (k_1 + k_2 + k_3 + k_4)^2 \sim 0 ~, ~~~~~~ t_{13} =(k_1 + k _3)^2 \sim 0
\auto\label{coff}
$$
translates into a rapidity cut-off on the relative rapidity of the
$(t_5,t_1,t_2)$ and $(t_6,t_3,t_4)$ vertices that determines the overall
normalization when we combine the reggeon diagram of Fig.~6.4 with that of
Fig.~6.3.

When (\ref{coff}) is satisfied we can write
$$
\tilde{G}_{\til{\alpha}}(j,t^-) ~\sim ~\tilde{C}~{ R_L^2~R_R \over (n_5 + n_6 -
\alpha_{13}^* -\alpha_{24}^*)}
\auto\label{rpf4}
$$
where $R_L$ is the triple Regge vertex,
$\alpha_{ij}~=~\alpha\bigl((k_i+k_j)^2\bigr) $,
and $\tilde{C}$ depends on the transverse momentum scale.
Combining (\ref{rpf4}) with (\ref{rpf3}) and inserting in (\ref{dis4r}) we
obtain, instead of the helicity integral (\ref{k0p}),
$$
\eqalign{~~~~~ & \int { dn_5dn_6 \over
(j-n_5-n_6 +1)(n_5 -\alpha_{12})(n_6 -\alpha_{34})
(n_5 + n_6 -
\alpha_{13}^* -\alpha_{24}^*)}\cr
&\to
{1 \over
(j-\alpha_{12} -\alpha_{34} +1)(j-\alpha_{13}^* -\alpha_{24}^* +1)}}
\auto\label{k0p2}
$$
Returning to (\ref{dis4r}) we obtain
$$
\eqalign{ \delta_{q^2}\biggl\{a^+(j,q^2)\biggr\}~\sim~& {g^4~N^2 \over 2}
\delta_{q^2}\Bigg\{J_3(q^2)\Biggl[ {1 \over
(j-\alpha_{12} -\alpha_{34} +1)(j-\alpha_{13}^* -\alpha_{24}^* +1)}\Biggr]
\Biggr\} }
\auto\label{dis4rb}
$$
where the color factor $N^2/ 2$ is obtained from Fig.~6.5.
\begin{center}
\leavevmode
\epsfxsize=4in
\epsffile{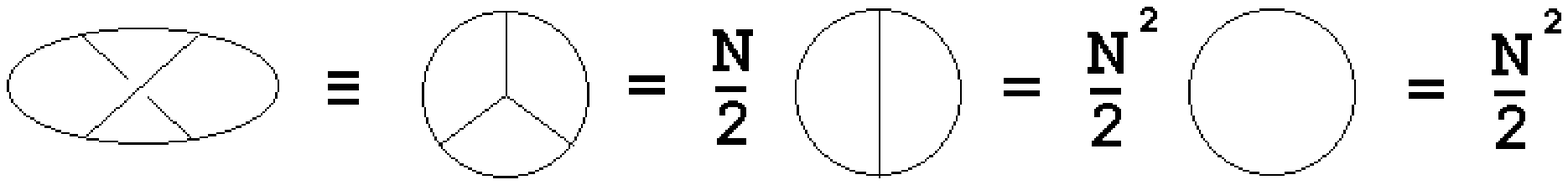}

Fig.~6.5 Color factor for the box diagram.
\end{center}
Clearly (\ref{dis4rb}) is the four-particle discontinuity of the reggeon
diagram Fig.~5.4 with $K^{(4)}_4$ as the reggeon interaction.

There are some important general points related to this last derivation
which we postpone discussion of until after we have discussed the remaining
terms in $K^{(4)}_{2,2}$. These involve reggeon diagrams of the form of
Fig.~6.6 that contain 1-3 reggeon interactions, either via an intermediate
reggeon or as a direct coupling, and are naturally defined in the coupling
scheme of Fig.~6.7.

\begin{center}
\leavevmode
\epsfxsize=2.7in
\epsffile{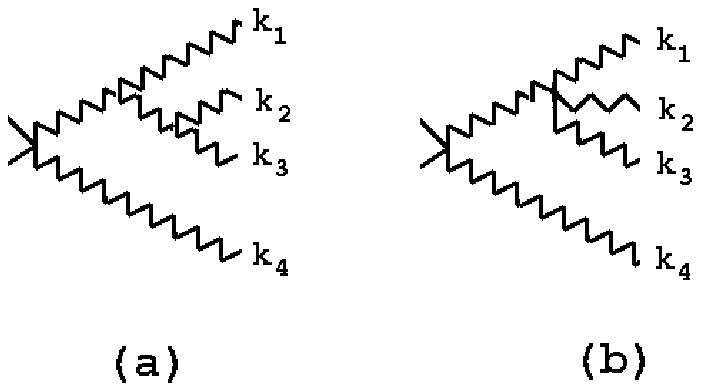}

Fig.~6.6 (a) 1-3 reggeon interaction via an intermediate reggeon (b) an
elementary 1-3 reggeon interaction.
\end{center}

\begin{center}
\leavevmode
\epsfxsize=3in
\epsffile{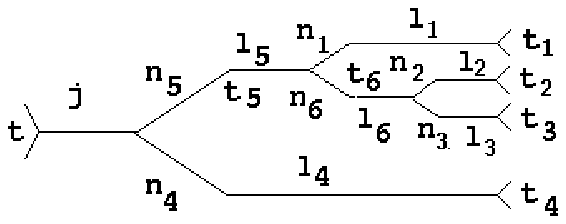}

Fig.~6.7 Alternative coupling scheme
\end{center}
A nonsense zero will automatically prevent the intermediate reggeon in
Fig.~6.6(a) from participating in a Regge cut. The Ward identity constraints
then determine that this diagram gives a 1-3 reggeon interaction of the form
$$
r_{13}(k_1,k_2,k_3)~\sim~{k_1.(k_2 +k_3)k_2.k_3 \over (k_2 +k_3)^2}
\auto\label{13int}
$$
However, in the limit $k_1^2, k_2^2, k_3^2 \to 0$ this reduces to $k_1.k_2 +
k_1.k_3$ and summing over such terms we simply obtain
$$
r_{13}(k_1,k_2,k_3)~\sim~(k_1+k_2+k_3)^2~=~k^2 ~.
\auto\label{r13}
$$
Therefore for the purpose of studying threshold contributions it suffices to
take $r_{13} \sim k^2$, as in Fig.~6.1(1), (2) and (3). In this case the
diagram of Fig.~6.6(a) should not be distinguished from that of Fig.~6.6(b).

To obtain the diagrams of Fig.~6.1(2) we need to consider the product of a
diagram of the form of Fig.~6.3 with the conjugate of Fig.~6.6(b). Again the
transformation from the coupling scheme of Fig.~6.7 to that of Fig.~6.2 is
simple only if we impose an additional kinematic constraint, such as
$(k_1+k_2+k_3)^2 \sim 0$, which requires that all the momenta participating
in the threshold are (close to) parallel. This constraint also leads to
an uncertainty in the overall normalization. However, after it is imposed,
it is straightforward to essentially repeat (\ref{rpf4}) - (\ref{dis4rb})
and obtain the first diagram of Fig.~6.1(2) as the reggeon interaction. The
second diagram is obviously obtained by the conjugate analysis.

Fig.~6.1(1) is a product of diagrams of the form of Fig.~6.6(b). Since these
diagrams are described by the same partial-wave it is, at first sight,
simple to derive as an immediate generalization
of (\ref{rpf})-(\ref{dis3r2}). Similarly Fig.~6.1(3) is a non-diagonal
product of diagrams of the form of Fig.~6.6(b). As a result, the derivation is
an immediate generalization of that of $K_2$ in Section 5.
There are, however, some important qualifications that we must make. The 1-3
reggeon coupling appearing in Fig.~6.6(b) can be generated by intermediate
$t$-channel states that are not nonsense states. Consequently we can not
argue immediately that the integrals appearing in Figs.6.1(2) and (4) are
transverse momentum integrals. The following discussion shows, nevertheless
that if these diagrams appear at all in the scale-invariant approximation
then a transverse momentum integral must be an infra-red approximation.

Because Fig.~6.1(2) involves the reggeon diagram of Fig.~6.3,
its' derivation is complete (for $q^2 = 0$) for the leading threshold behavior
given by the ``bubble diagram'', which is then necessarily
a transverse momentum diagram. As we have discussed in \cite{cw}, the 1-3
reggeon coupling appearing in Figs.~6.1(1), (2) and (3) is determined by a
Ward identity constraint involving Figs.~6.1(2) and (3). This constraint is
illustrated in Fig.~6.8 (a dashed line indicates zero transverse momentum)
\begin{center}
\leavevmode
\epsfxsize=3in
\epsffile{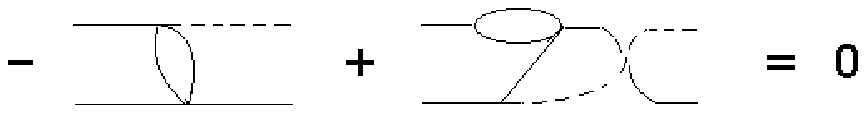}

Fig.~6.8 The Ward identity constraint for $K^{4n}$.
\end{center}
and has the form
$$
r_{13}^2~\sim ~r_{13}r_{ijk}
\auto\label{wdc}
$$
giving both zero and non-zero solutions for $r_{13}$. If we
assume the non-zero solution is the correct one, then this constraint
implies that Fig.~6.1(3) must be present, in addition to Fig.~6.1(2), as a
transverse momentum diagram. Infra-red finiteness is then satisfied provided
the full box diagram of Fig.~6.1(4) is present. (That is the part of the
box diagram containing thresholds in $(k_1-k_3)^2$ and $(k_1-k_4)^2$, which are
not accessible within the $t$-channel unitarity integral must also be
present.) The presence of Fig.~6.1(1),
which can not be distinguished\cite{cw} from Fig.~6.1(0) in the forward
direction, is then required by infra-red finiteness of the kernel after
integration\cite{ker,cw}.

It is clear that in discussing the diagrams of Fig.~6.1 we have had, in
each case, to impose a leading threshold behavior constraint {\it in addition
to imposing $q^2 \to 0$.} This means that our discussion is restricted to the
forward kernel and to the leading infra-red behaviour in the
additional transverse momentum variables, producing related uncertainties
in overall normalization. In particular, the derivation of $K^{(4)}_4$
applies to the leading threshold behavior as $(k_1+k_3)^2 \to 0$ and by
analogy the leading behavior when $(k_2+k_4)^2 \to 0$, $(k_1+k_2)^2 \to 0$
and $(k_3+k_4)^2 \to 0$. In \cite{cw} we argued that at $q^2 = 0$,
these thresholds can be extracted from the box diagram and we can write
\beq
K_{2,2}^{(4n)}= (K_{BFKL}/2)^2 + {\cal K}_2
\label{complete}
\eeq
where ${\cal K}_2$ contains the thresholds and is a separate
infra-red finite kernel. We also derived the eigenvalue spectrum of
${\cal K}_2$ and showed that it satisfies the crucial property of
{\it holomorphic factorization.}

The analysis of this Section can be summarized as showing that the forward
kernel does indeed contain the two terms present in (\ref{complete}) but we
can not determine their relative normalization. The fundamental new result
is clearly that the ${\cal K}_2$ component can be unambiguously derived from
unitarity. The infra-red finiteness and Ward identity constraints imply that
all the remaining parts of $K^{(4n)}_{2,2}$ can be uniquely written as
$(K_{BFKL})^2$. There is also a further very important property which
distiguishes ${\cal K}_2$ from To properly explain this requires the full
asymptotic dispersion theory of \cite{arw1}. We can briefly
summarize the essential point as follows.

As elaborated in \cite{arw1}, the different multiple discontinuities of a
multiparticle amplitude have different continuations to complex angular
momenta and helicities. We can embed $K^{(4)}_4$, as a reggeon interaction,
in an eight-point amplitude as illustrated in Figs.~6.9 and 6.10. There are
then two classes of multiple discontinuities that have to be considered in
defining the $j$-plane continuations that we have utilised. An example of
the first class is illustrated in Fig.~6.9.
\begin{center}

\leavevmode
\epsfxsize=5in
\epsffile{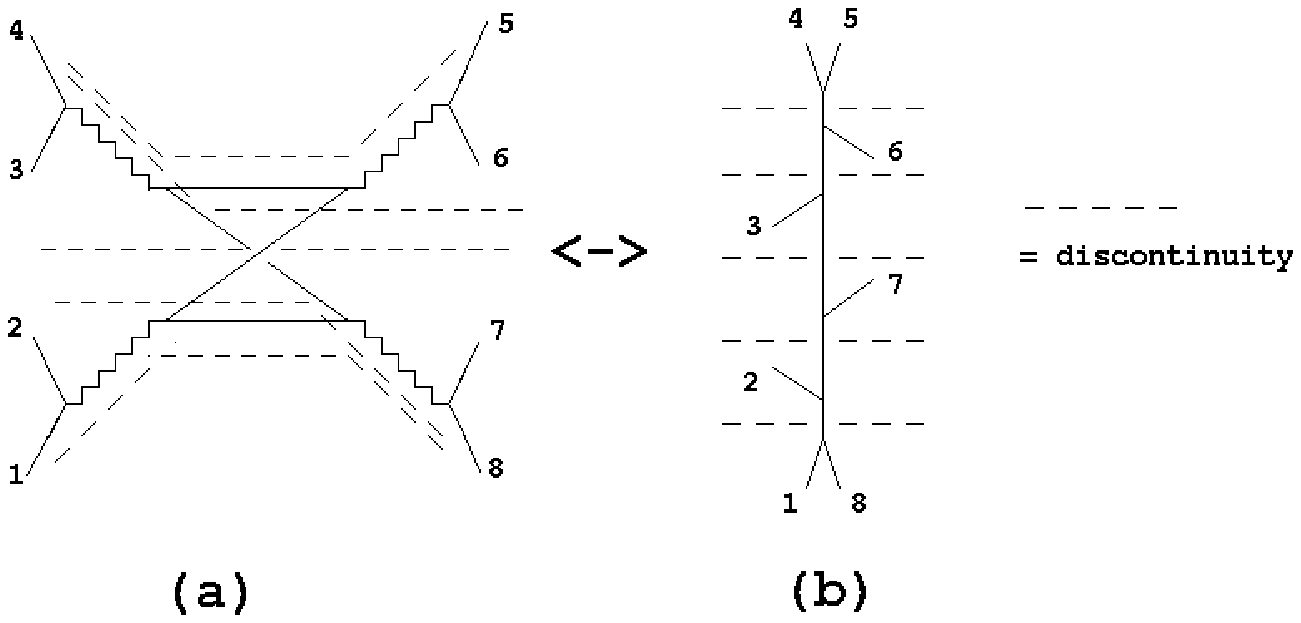}

Fig.~6.9 (a) Multiple discontinuities of the eight-point amplitude (b) the
corresponding tree graph.
\end{center}
Other examples in the same class would interchange the reggeons relative to
the discontinuities taken. An example of the second class is illustrated in
Fig.~6.10.
\begin{center}

\leavevmode
\epsfxsize=5in
\epsffile{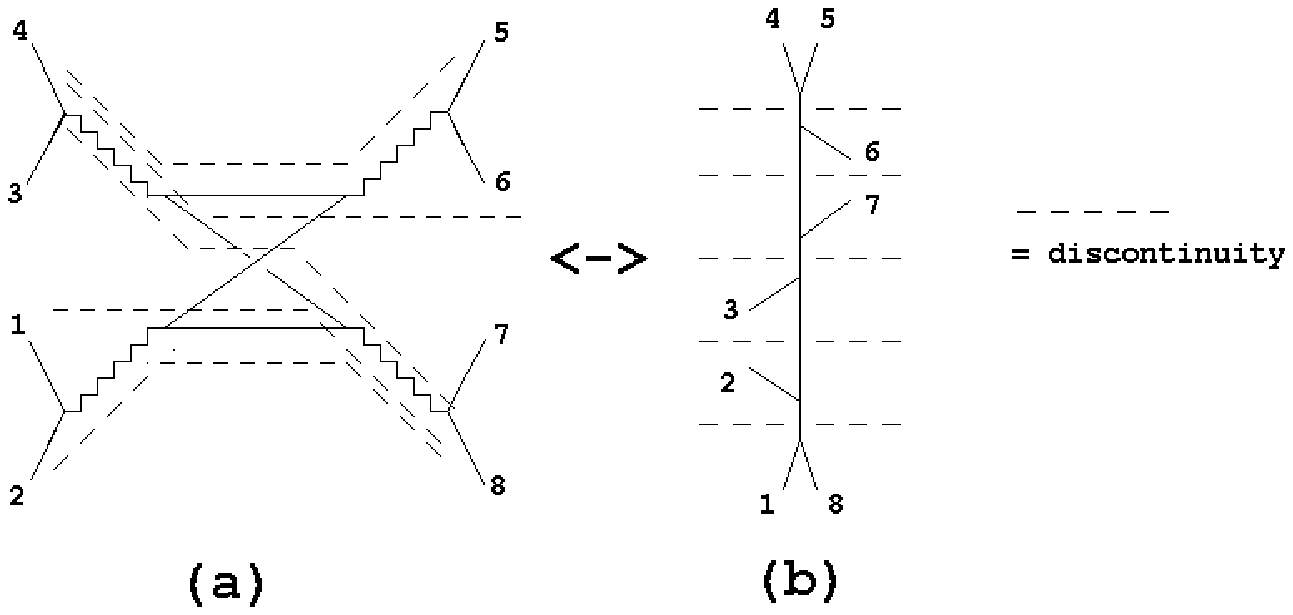}

Fig.~6.9 (a) Alternative discontinuities of the eight-point amplitude (b) the
corresponding tree graph
\end{center}

As in the examples we have shown, the two classes of multiple discontinuities
differ in general only by one discontinuity. Fig.~6.9 contains the
$s_{3456}$ discontinuity, whereas Fig.~6.10 contains instead the $s_{4567}$
discontinuity. The $s_{3456}$ discontinuity necessarily involves also taking
a discontinuity of the box diagram (of the $(k_1 -k_3)^2$ form) whereas the
$s_{4567}$ discontinuity does not. All the connected diagrams of
$(K_{BFKL})^2$ contain the necessary discontinuity (as do those of the
leading-order $K_{BFKL}$.) On the other hand, the thresholds of ${\cal K}_2$
are compatible only with the $s_{4567}$ discontinuity. It follows that the
two terms in (\ref{complete}) will be separated by the process of taking the
multiple discontinuities. $(K_{BFKL})^2$ will appear in Fig.~6.9
whereas ${\cal K}_2$ will appear in Fig.~6.10.

We use the partial-wave projection of Fig.~4.3 and consider the continuation to
complex $j,~n_1,~n_2~n'_1,~n'_2$. Corresponding to their distinction with
respect to one discontinuity, the multiple discontinuities of Figs.~6.9 and
6.10 differ in one feature of the analytic continuations made. For example,
in both cases a continuation is made from
$$
n_2~>~0,~~n'_1~>~0~,~~n'_2~>~n_2~,~~n_1 +n_2~>~n'_1 +n'_2
\auto\label{cont1}
$$
whereas, for Fig.~6.9 the final continuation is made from
$$
j~>~n_1~+~n_2
\auto\label{cont2}
$$
while for Fig.~6.10 it is made from
$$
{}~~~j~< ~n_1+n'_2
\auto\label{cont3}
$$
As a result a distinct analytically continued amplitude is defined in each
case.

Although both partial-wave amplitudes appear similarly in the BFKL kernel,
they are distinguished in the more general framework of multi-Regge theory.
Since $K_{BFKL}$ contributes only to Fig.~6.9, it follows that the
separation in (\ref{complete}) distinguishes a leading-order contribution to
a new partial-wave amplitude, i.e. ${\cal K}_2$, from a non-leading
contribution to the partial-wave amplitude that also contains the
leading-order kernel. Not surprisingly the non-leading contribution suffers
most from scale ambiguities.

\mainhead{7. CONCLUSIONS}

We have demonstrated that the direct analysis of $t$-channel unitarity
is able to give a firm foundation to the results of \cite{ker}. Not
surprisingly the limitations of the results are also apparent. These include
uncertainties in normalization due to scale-dependence and infra-red
kinematical constraints. Our non-leading results are essentially summarized
by writing for the full BFKL kernel $K_{2,2}(q,k,k')$
$$
K_{2,2}(q,k,k')~\centerunder{$\longrightarrow$}{\raisebox{-3mm}
{$\scriptstyle q^2, k^2, k'^2 \to 0 $}}~~ g^2 K_{BFKL} +
O(g^4) (K_{BFKL})^2 + O(g^4) {\cal K}_2 ~
\auto\label{suma}
$$
indicating that both the overall normalization and
the normalization of ${\cal K}_2$ relative to $(K_{BFKL})^2$ are not
determined. Results obtained by Kirschner\cite{kir} from the multi-Regge
effective lagrangian are completely consistent with (\ref{sum}), although the
separate significance of ${\cal K}_2$ is not apparent in the $s$-channel
formalism. A fundamental new result obtained in this paper is that ${\cal K}_2$
can be unambiguously derived from $t$-channel unitarity as the leading-order
contribution of a new (analytically continued) multiparticle partial-wave
amplitude. We had previously identified ${\cal K}_2$  only as a separately
infra-red finite component of the kernel. The holomorphic factorization
properties of its' spectrum are strongly suggestive that the
full (non-forward) leading-order form of this amplitude will be conformally
invariant. Indeed in a separate paper\cite{cpw} we have constructed a
candidate for this amplitude using the Ward identity constraints.

It is clear that if a reggeon interaction is unambiguously derivable from
$t$-channel unitarity, it will necessarily be scale-invariant and presumably
must be the leading-order infra-red approximation to some well-defined
partial-wave amplitude. The extrapolation away from the infra-red region
will then satisfy Ward identity constraints which, we conjecture,
necessarily lead to conformal invariance. From the multi-Regge theory of
\cite{arw1} we know there exists a vast array of distinct partial-wave
amplitudes describing Regge limits of dispersion-relation defined components
of multiparticle amplitudes. The BFKL kernel, the triple Regge
kernel\cite{bw,ker}, and the new ${\cal K}_2$ kernel we have derived are
amongst the simplest examples. Our results and those of \cite{blw}, when added
to previous results on the BFKL kernel, are consistent with the conjecture
that all such amplitudes have a leading-order conformally invariant
approximation. The physical significance of this approximation remains to be
determined.

In \cite{uni} we have also outlined a program whereby the scale-dependence of
non-leading reggeon amplitudes might be studied via the Ward identity
constraints. We leave to a future study the possibility that the formalism of
this paper can be extended in this direction.

\centerline{\bf Acknowledgements}

We are grateful to J. Bartels, R. Kirschner and L. Lipatov for valuable
discussions of this work.

\end{document}